\documentclass[12pt]{article}
\usepackage{amsmath,amssymb,epsfig}
\makeatletter \@addtoreset{equation}{section} \makeatother
\renewcommand{\theequation}{\thesection.\arabic{equation}}
\newcommand{\ba}{\begin{array}}
\newcommand{\ea}{\end{array}}
\newcommand{\beq}{\begin{equation}}
\newcommand{\eeq}{\end{equation}}
\newcommand{\bea}{\begin{eqnarray}}
\newcommand{\eea}{\end{eqnarray}}

\def\bce{\begin{center}}
\def\ece{\end{center}}

\def\nonu{\nonumber}

\def\be{\beta}

\newcommand{\tr}{\mbox{Tr}}

\def\eps6{{\displaystyle \mathop{\epsilon}^{6}}{}}

\def\nab6{{\displaystyle \mathop{\nabla}^{6}}{}}

\def\0{{\sst{(0)}}}
\def\1{{\sst{(1)}}}
\def\2{{\sst{(2)}}}
\def\3{{\sst{(3)}}}
\def\4{{\sst{(4)}}}
\def\5{{\sst{(5)}}}
\def\6{{\sst{(6)}}}
\def\7{{\sst{(7)}}}
\def\8{{\sst{(8)}}}

\def\ba{\begin{array}}
\def\ea{\end{array}}
\def\beq{\begin{equation}}
\def\eeq{\end{equation}}
\def\be{\begin{equation}}
\def\ee{\end{equation}}

\def\tr{\mathop{\rm tr}}

\def\eps{\epsilon}

\def\ba{\begin{array}}
\def\ea{\end{array}}
\def\beq{\begin{equation}}
\def\eeq{\end{equation}}
\def\be{\begin{equation}}
\def\ee{\end{equation}}

\def\tr{\mathop{\rm tr}}

\def\eps{\epsilon}

\newcommand{\bean}{\begin{eqnarray*}}
\newcommand{\eean}{\end{eqnarray*}}

\begin{document}
\thispagestyle{empty} \addtocounter{page}{-1}
   \begin{flushright}
KIAS-P08019 \\
\end{flushright}

\vspace*{1.3cm}

\centerline{ \Large \bf  Meta-Stable Brane Configurations,  }
\vspace{.3cm} 
\centerline{ \Large \bf  Multiple NS5-Branes, and Rotated D6-Branes } 
\vspace*{1.5cm}
\centerline{{\bf Changhyun Ahn} 
} 
\vspace*{1.0cm} 
\centerline{\it 
Department of Physics, Kyungpook National University, Taegu
702-701, Korea} 
\vspace*{0.8cm} 
\centerline{\tt ahn@knu.ac.kr} 
\vskip2cm

\centerline{\bf Abstract}
\vspace*{0.5cm}

We construct the type IIA
nonsupersymmetric meta-stable brane
configurations corresponding to 
the various ${\cal N}=1$ supersymmetric gauge theories.
The D6-branes are both displaced and rotated 
where these deformations 
are described as the mass term and quartic term
for the fundamental flavors respectively.
The multiplicity of the NS5-branes occurs in the
superpotential order for adjoint, symmetric, or bifundamental matters.  
A rich pattern of nonsupersymmetric meta-stable states as well
as the supersymmetric stable states is found.

\baselineskip=18pt
\newpage
\renewcommand{\theequation}
{\arabic{section}\mbox{.}\arabic{equation}}

\section{Introduction}

The dynamical supersymmetry breaking in meta-stable vacua \cite{ISS,IS} 
occurs
in the ${\cal N}=1$ supersymmetric gauge theory with massive fundamental 
flavors and 
the corresponding type IIA meta-stable brane constructions  
have been studied in \cite{OO1,FGU,BGHSS}.  
Other type IIA nonsupersymmetric meta-stable 
brane configuration has been constructed by considering  
the additional  quartic
term for the quarks in the 
electric superpotential besides the mass term \cite{GK0710-1,GK0710}.
This extra deformation in the ${\cal N}=1$ supersymmetric gauge theory 
corresponds to the rotation of D6-branes 
along the (45)-(89) directions in type IIA string theory. 
The nonsupersymmetric ground states arise only after 
the gravitational attraction of NS5-brane \cite{GK} is considered.

Branes can be used to describe the dynamics of a wide variety of 
${\cal N}=1$ supersymmetric gauge theories with different matter
contents and superpotential \cite{GK98}. 
It is very important to control the gauge singlet meson fields 
in appropriate way in order to find out the new meta-stable states.
In general, one can consider the electric brane configuration of
$k$ coincident left NS5-branes(012345) connected by $N_c$ D4-branes to 
$k'$ coincident right 
NS5'-branes(012389), with $N_f$ D6-branes(0123789) located
between the left NS5- and right NS5'-branes \cite{EGKRS}.  
The creation of D4-branes during the Seiberg dual is related to the
gauge singlets in the magnetic brane configuration.
Then there exist $k$ magnetic meson fields because 
each NS5-brane in the above brane configuration produces 
each meson field when it is crossing the D6-branes.

One can try to find out the new meta-stable brane configuration for 
a single meson field by taking 
a single left NS5-brane but multiple $k'$ right NS5'-branes rather than the
general case with $k$ left NS5-branes and $k'$ right NS5'-branes where the
corresponding dual gauge theory is not known so far.   
This question is also raised in \cite{Ahn06} where the meta-stable
brane configuration when $k=2$ and $k'=1$ was studied.
In this paper, 
we find the various meta-stable brane configurations 
for the different gauge theories with matters and superpotential,
along the lines of \cite{Ahn07-11,Ahn08-1,Ahn08-1two,Ahn08-2}.
All of these examples contain a set of a single NS5-brane and multiple 
NS5'-branes in their type IIA brane configurations.
The single magnetic meson field is realized by the product of quark
and anti-quark in the electric gauge theory 
in the presence of D6-branes with 
quartic superpotential for the fundamentals. 
On the other hand, 
when there are no D6-branes, it is realized by the product of 
bifundamentals with higher order superpotential.

In section 2, we review the type IIA brane configuration corresponding
to the ${\cal N}=1$ $SU(N_c)$ gauge theory 
with two adjoints and fundamentals and deform this theory 
by adding both the mass term
and the quartic term for the fundamentals. 
Then we describe the dual 
${\cal N}=1$ $SU(\widetilde{N}_c)$ gauge 
theory with corresponding dual
matter as well as a gauge singlet. 
We discuss the nonsupersymmetric meta-stable
minimum  and present 
the corresponding 
intersecting brane configuration of type IIA string
theory.
In section 3, we apply the method of section 2 to the ${\cal N}=1$ $Sp(N_c)$ 
gauge theory 
with two adjoints and fundamentals.

In section 4, we review the type IIA brane configuration corresponding
to the ${\cal N}=1$ $SU(N_c)$ gauge theory 
with symmetric tensor and fundamentals and deform this theory 
by adding both the mass term
and the quartic term for the fundamentals. 
Then we study the dual 
${\cal N}=1$ $SU(\widetilde{N}_c)$ gauge 
theory with corresponding dual
matters. 
We present the nonsupersymmetric meta-stable
minimum   and 
the corresponding 
intersecting brane configuration of type IIA string
theory.

In section 5, we review the type IIA brane configuration corresponding
to the ${\cal N}=1$ $SU(N_c) \times SU(N_c')$ gauge theory 
with bifundamentals and fundamentals and deform this theory 
by adding both the mass terms
and the quartic terms for each fundamentals. 
Then we describe the dual 
${\cal N}=1$ $SU(\widetilde{N}_c) \times SU(N_c')$ gauge 
theory with corresponding dual
matters. 
We discuss the nonsupersymmetric meta-stable
minimum  and  
the corresponding 
intersecting brane configuration.
In section 6, we use the method of section 5 to
the ${\cal N}=1$ $Sp(N_c) \times SO(2N_c')$ gauge theory 
with bifundamentals, vectors and fundamentals. 
In section 7, we apply the method of section 5 to
the ${\cal N}=1$ $SU(N_c) \times SO(2N_c')$ gauge theory 
with bifundamentals, vectors and fundamentals.

In section 8, we review the type IIA brane configuration corresponding
to the ${\cal N}=1$ $SU(N_c) \times SU(N_c')$ gauge theory 
with symmetric tensor and bifundamentals  and deform this theory 
by adding both the mass term
and the higher order term for bifundamentals. 
Then we describe the dual 
${\cal N}=1$ $SU(\widetilde{N}_c) \times SU(N_c')$ gauge 
theory with corresponding dual
matters. 
We discuss the nonsupersymmetric meta-stable
minimum  and  
the corresponding 
intersecting brane configuration.
In section 9, we use the method of section 8 to
the ${\cal N}=1$ $SU(N_c) \times SU(N_c')$ gauge theory 
with antisymmetric tensor, eight fundamentals and bifundamentals.

In section 10, we review the type IIA brane configuration corresponding
to the ${\cal N}=1$ $SU(N_c) \times SU(N_c')$ gauge theory 
with symmetric tensor, bifundamentals and fundamentals and deform this theory 
by adding both the mass terms
and the quartic terms for each fundamentals. 
Then we describe the dual 
${\cal N}=1$ $SU(\widetilde{N}_c) \times SU(N_c')$ gauge 
theory or 
${\cal N}=1$ $SU(N_c) \times SU(\widetilde{N}_c')$ gauge 
theory
with corresponding dual
matters. 
We study the nonsupersymmetric meta-stable
minimum  and  
the corresponding 
intersecting brane configurations.
In section 11, we also describe
the ${\cal N}=1$ $SU(N_c) \times SU(N_c')$ gauge theory 
with antisymmetric tensor, bifundamentals and fundamentals.

In section 12,  we make some comments for the future directions.  

\section{$SU(N_c)$ with two adj. and $N_f$-fund. }

\subsection{Electric theory}

The type IIA supersymmetric electric
brane configuration \cite{EGKRS} corresponding to 
${\cal N}=1$ $SU(N_c)$ gauge theory  with 
two adjoint fields $\Phi$, $\Phi'$ and 
$N_f$-fundamental flavors $Q, \widetilde{Q}$
can be described as follows: one NS5-brane(012345), $k'$
NS5'-branes(012389), 
$N_c$
D4-branes(01236), and $N_f$ D6-branes(0123789). 
Let us introduce two complex coordinates \cite{GK98}
\bea
v \equiv x^4 + i x^5, \qquad w \equiv x^8 + i x^9.
\nonu
\eea
Since we consider a single NS5-brane, the adjoint field $\Phi$ is massive and
can be integrated out. For large coupling in front of the
quadratic $\Phi$ term, 
there is no $\Phi$-dependence in the superpotential.
The $N_c$-color D4-branes are suspended between 
the left NS5-brane and the right NS5'-branes and similarly the $N_f$ D6-branes 
are located between the left NS5-brane and the right NS5'-branes.
The
$N_f$-fundamental flavors $Q, \widetilde{Q}$  are strings
stretching between $N_f$ D6-branes and $N_c$-color D4-branes while 
the adjoint field $\Phi'$
is related to the fluctuations of $N_c$-color D4-branes in $v$ direction. 

Let us deform this theory
by adding the mass term 
and the quartic term for fundamental quarks $Q, \widetilde{Q}$ 
in order to find the new
nonsupersymmetric meta-stable states.
The former can be achieved by displacing the D6-branes along $+v$
direction leading to their coordinates $v = + v_{D6}$ \cite{GK98} 
while the latter can be obtained by rotating the D6-branes
\cite{GK0710-1} 
by an angle 
$-\theta$ in $(w,v)$-plane and we denote those rotated D6-branes 
by $D6_{-\theta}$-branes. 
Then, in the electric gauge theory, the deformed superpotential is
given by
\bea
W_{elec} = \left[ \frac{g_{\Phi'}}{2} \tr {\Phi'}^{k'+1}+  Q \Phi'
  \widetilde{Q} 
\right]  - m \tr Q
\widetilde{Q} +
\frac{\alpha}{2} \tr (Q \widetilde{Q})^2,
\qquad \alpha = \frac{\tan \theta}{\Lambda}, \qquad 
m = \frac{v_{D6_{-\theta}}}{2\pi \ell_s^2}
\label{electricsuperpotential}
\eea 
where $\Lambda$ is related to the scales of the electric and magnetic 
theories and $+v_{D6_{-\theta}}$ is 
the $v$ coordinate of the center of coincident rotated 
$D6_{-\theta}$-branes.  
When $k'=1$, this theory reduces to the one in \cite{GK0710-1,GK0710}
because the first two terms of (\ref{electricsuperpotential}) contribute to
the additional quartic term for 
the quarks $Q,\widetilde{Q}$ and we are left with the last two terms
in (\ref{electricsuperpotential}).
Therefore, we focus on the nontrivial case with the number of
NS5'-branes $k' \geq 2$.

Let us summarize the ${\cal N}=1$ supersymmetric electric brane
configuration 
\footnote{Without the quartic term($\alpha=0$), the theory
with $k'=N_c-1$ obtained 
by resolving the superpotential(i.e., there exist also the lower
order terms by splitting the $k'$ NS5'-branes in $v$ direction) 
was studied in \cite{dO} some time ago in the supersymmetric brane
configuration
and this theory
without D6-branes when $k' \geq N_c+1$(multitrace interactions) 
was also studied in the context
of supersymmetry breaking vacua from M5-branes recently \cite{MOY}. }
with superpotential 
(\ref{electricsuperpotential}) 
in type IIA string theory as follows and draw this in
Figure 1:

$\bullet$ One left NS5-brane in (012345) directions with $w=0$

$\bullet$ $k'$ right NS5'-branes in (012389) directions with $v=0$

$\bullet$ $N_c$-color D4-branes in (01236) directions with $v=0=w$

$\bullet$ $N_f$ $D6_{-\theta}$-branes in (01237)
directions and
two other directions in $(v,w)$-plane  

\begin{figure}[ht]
   \epsfxsize=1.0in 
\centerline{\epsffile{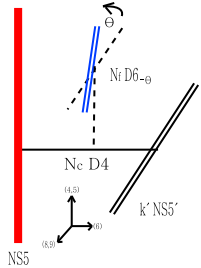}}
   \caption[FIG. \arabic{figure}.]{ 
The  ${\cal N}=1$ supersymmetric 
electric brane configuration for the gauge group $SU(N_c)$ 
with two adjoints and fundamentals $Q, \widetilde{Q}$. 
Note that there are $k'$ NS5'-branes and the fluctuations of 
$N_c$ color D4-branes
in $v$ direction correspond to the adjoint field.
A 
``rotation'' of $N_f$ D6-branes in $(w,v)$-plane, which become
$D6_{-\theta}$-branes, 
corresponds to 
a quartic term for the fundamentals while 
a ``displacement'' of $N_f$ D6-branes in $+v$ direction corresponds to a
mass term for the fundamentals.
}
\end{figure}

\subsection{Magnetic theory}

Let us move the left NS5-brane to the right all the way past the right 
NS5'-branes and one arrives at the Figure 2A.
Note that there exists a creation of $N_f$-flavor D4-branes
connecting $N_f$ $D6_{-\theta}$-branes and $k'$ NS5'-branes.
Recall that the $N_f$ $D6_{-\theta}$-branes are not parallel to the 
NS5-brane in Figure 1 unless $\theta =\frac{\pi}{2}$.
The linking number \cite{HW} of the NS5-brane from Figure 2A
is 
$
l_m = \frac{N_f}{2} -\widetilde{N}_c$.
On the other hand, the linking number of the NS5-brane from Figure 1
is
$
l_e = -\frac{N_f}{2} + N_c $.
From the equality between these two relations, one obtains
the number of colors of dual magnetic theory \cite{EGKRS}
\bea
\widetilde{N}_c =N_f-N_c.
\nonu
\eea
Note that the multiplicity $k'$ of NS5'-branes does not arise in this 
computation because the creation of flavor D4-branes appears via the
dual process between
the NS5-brane and $N_f$ $D6_{-\theta}$-branes 
which do not depend on $k'$. 

\begin{figure}[ht]
   \epsfxsize=3.0in 
\centerline{\epsffile{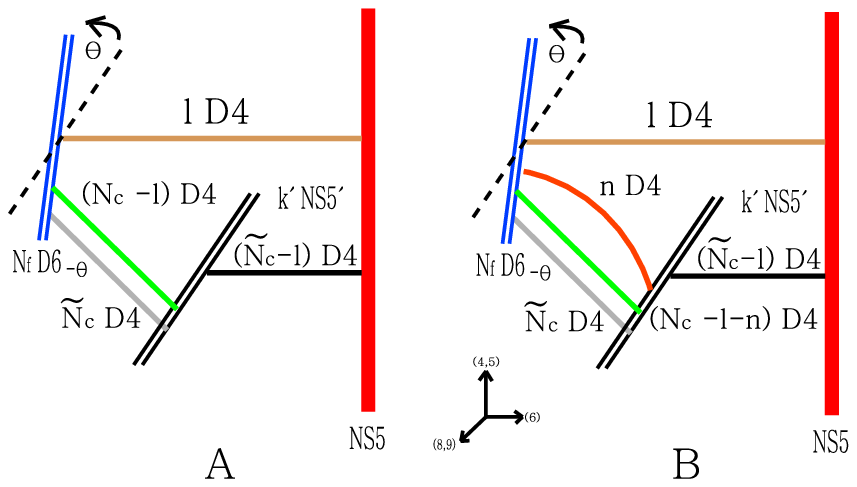}}
   \caption[FIG. \arabic{figure}.]{ 
The  ${\cal N}=1$ supersymmetric
magnetic brane configuration corresponding to Figure 1 with 
a misalignment between D4-branes when the gravitational potential of
the NS5-brane is ignored(2A) and nonsupersymmetric brane configuration
when  the gravitational potential of
the NS5-brane is considered(2B). 
The $N_f$ flavor D4-branes connecting between
$D6_{-\theta}$-branes and NS5'-branes are splitting into $ 
\widetilde{N}_c$-, $(N_c-l)$- and
$l$- D4-branes(2A). Further $n$- D4-branes among 
$(N_c-l)$- D4-branes are moved to the NS5-brane(2B). 
}
\end{figure}

The low energy theory on the dual color D4-branes 
has $SU(\widetilde{N}_c)$ gauge group and an adjoint field 
$\phi'$ coming from 4-4 strings connecting the dual color D4-branes 
and 
$N_f$-fundamental dual quarks $q, \widetilde{q}$
coming from 4-4 strings connecting between the color D4-branes and
flavor D4-branes.
Moreover, a single magnetic meson field $M \equiv Q \widetilde{Q}$
is $N_f \times N_f$ matrix and comes from 
4-4 strings of flavor D4-branes.
Then the dual magnetic superpotential is given by  
\bea
W_{mag} = \left[ \frac{g_{\phi'}}{2} \tr {\phi'}^{k'+1}+  
q \phi' \widetilde{q} +
\frac{1}{\Lambda} M q \widetilde{q} \right] 
+ \frac{\alpha}{2} \tr M^2 - m M.
\label{superpo}
\eea 
The undeformed expression, the first three terms, 
is already found in \cite{EGKRS}.
The case with $k'=1$ leads to the one in \cite{GK0710-1}.
In order to obtain the supersymmetric vacua, 
one computes the F-term equations for 
the superpotential (\ref{superpo}):
\bea
 \frac{1}{\Lambda} M q + q \phi'& = & 0, 
\qquad \phi' \widetilde{q} +\frac{1}{\Lambda} \widetilde{q} M =0, 
\nonu \\
\frac{1}{2} g_{\phi'}(k'+1) {\phi'}^{k'} + \widetilde{q} q & = & 0,
\qquad \frac{1}{\Lambda} q  \widetilde{q}  =  m -\alpha M. 
\label{Fterm}
\eea

We choose the adjoint field $\phi'$ to be diagonal, i.e., 
$\phi' = \mbox{diag} (\phi'_1, \cdots, \phi'_{\widetilde{N}_c})$.
Then the first two equations of (\ref{Fterm}) imply that 
the upper left $\widetilde{N}_c \times \widetilde{N}_c$ block 
of $\frac{M}{\Lambda}$ is given
by $  \mbox{diag} (-\phi'_1, \cdots, -\phi'_{\widetilde{N}_c})$.
By substituting this into the last equation of (\ref{Fterm}), one
obtains 
the upper left $\widetilde{N}_c \times \widetilde{N}_c$ block 
of $\frac{1}{\Lambda} q \widetilde{q}$ is given
by $ \mbox{diag} ( m+\alpha \Lambda \phi'_1, \cdots, 
m+ \alpha \Lambda \phi'_{\widetilde{N}_c})$.
Moreover, the lower right $(N_f-\widetilde{N}_c) 
\times (N_f-\widetilde{N}_c)$ block 
of $M$ is given by $l$'s zero eigenvalues and 
$(N_f-\widetilde{N}_c-l)$'s eigenvalues
$\frac{m}{\alpha}$.
Then finally, 
the third equation of (\ref{Fterm}) gives rise to the expectation
value for the adjoint field $\phi'$ satisfying that 
\bea
\frac{1}{2} g_{\phi'}(k'+1) {\phi_j'}^{k'}=-\Lambda(m+\alpha \Lambda \phi_j')
\label{condition}
\eea
for $j=1,2, \cdots, \widetilde{N}_c$. 
For nonzero quark mass $m$, the expectation 
value $\phi_j'$ is not vanishing and $-\Lambda \phi_j' > \frac{m}{\alpha}$ for
the positiveness of left hand side of (\ref{condition}) when
$g_{\phi'}$ and $\phi_j'$ are real.

Let us first describe the nonsupersymmetric meta-stable states
and supersymmetric ones 
when all the  $N_f$ $D6_{-\theta}$-branes and $k'$ NS5'-branes
are coincident with each other.

$\bullet$ Coincident $N_f$ $D6_{-\theta}$-branes 
and $k'$ NS5'-branes(equal massive
flavors) 

One writes $N_f \times N_f$ matrix $M$  with $\widetilde{N}_c(=N_f-N_c)$
eigenvalues by the diagonal elements for 
$\phi'$, $l$'s eigenvalues by the zeros 
and $(N_c-l)$ eigenvalues
$\frac{m}{\alpha}$ 
as follows:
\bea
M = \left(
\begin{array}{ccc}
-\Lambda \phi' & 0  & 0  \\
0 & 0_l & 0
\\
0 & 0 &  \frac{m}{\alpha} {\bf 1}_{N_c-l}
\end{array}
\right), \qquad \mbox{with} \qquad 
\phi' = \mbox{diag} (\phi'_1, \cdots, \phi'_{\widetilde{N}_c}).
\label{M0}
\eea
In the brane configuration of Figure 2A, the $l$ of the
$N_f$-flavor D4-branes are connected with $l$-color
D4-branes
and the resulting $l$ D4-branes 
stretch from the $D6_{-\theta}$-branes to
the NS5-brane directly 
and the intersection point between the 
$l$ D4-branes and the $D6_{-\theta}$-branes is given by 
$(v, w)=(+v_{D6_{-\theta}}, 0)$.
This
corresponds to  exactly the $l$'s eigenvalues from 
zeros of 
$M$ in (\ref{M0}).
Now the remaining $(N_c-l)$-flavor D4-branes between 
the $D6_{-\theta}$-branes and 
the NS5'-branes correspond to the eigenvalues 
of $M$ in (\ref{M0}), i.e.,   
$\frac{m}{\alpha} {\bf 1}_{N_c-l}$.
The intersection point between the 
$(N_c-l)$ D4-branes and the NS5'-branes is given 
by $(v, w)=(0, +v_{D6_{-\theta}} \cot \theta)$ from trigonometric 
geometry \cite{GK0710-1}.
Finally, the remnant $\widetilde{N}_c$-flavor D4-branes
between 
the $D6_{-\theta}$-branes and 
the NS5'-branes 
correspond to the eigenvalues $-\Lambda \phi'$ in (\ref{M0}) with 
(\ref{condition}) providing the exact nonzero $w$ coordinates for
these flavor D4-branes. 

One also represents the vacuum expectation value for the quarks
as follows:
\bea
q  \widetilde{q} = \left(
\begin{array}{ccc}
\Lambda (m {\bf 1}_{\widetilde{N}_c} + \alpha \Lambda \phi')  & 0 & 0  \\
0 & \Lambda m {\bf 1}_l & 0 \\
0 & 0 & 0_{N_c-l} 
\end{array}
\right), \qquad \mbox{with} \qquad 
\phi' = \mbox{diag} (\phi'_1, \cdots, \phi'_{\widetilde{N}_c}).
\label{solqq}
\eea
In the $l$-th vacuum the gauge symmetry is broken to
$SU(\widetilde{N}_c-l)$ and when the supersymmetric vacuum is drawn in
Figure 2A with $l=0$, the gauge group 
$SU(\widetilde{N}_c)$ is unbroken. 
Then the supersymmetric 
ground state corresponds to the vacuum expectation values $M$ by
$-\Lambda \phi'$ which has $\widetilde{N}_c$'s eigenvalues and  
$\frac{m}{\alpha}$ with degeneracy $N_c$.

The theory has nonsupersymmetric meta-stable ground states
since there exists an attractive
gravitational interaction
between the flavor 
D4-branes and the NS5-brane from the DBI action 
\cite{GK}.
Let us rescale the meson field as
$M = h \Lambda \Phi $.
Then the magnetic superpotential (\ref{superpo}) 
can be rewritten in terms of $\Phi, q, \widetilde{q}$ and $\phi'$
\bea
W_{dual} = 
 h \Phi  q   \widetilde{q} 
 +  
\frac{ \mu_{\phi}}{2} h^2 \tr \Phi^2- h \mu^2 \tr \Phi +
\frac{g_{\phi'}}{2} \tr {\phi'}^{k'+1}+  q \phi' \widetilde{q}
\label{Dualsec2}
\eea
with 
$
\mu^2 = m \Lambda$ and  
$\mu_{\phi} = \alpha \Lambda^2$.

Now one splits 
the $(N_c-l) \times (N_c-l)$
block  at the lower right corner of $M$ and $q \widetilde{q}$ 
into blocks of 
size $n$ and $(N_c-l-n)$ and then 
(\ref{M0}) and (\ref{solqq}) are 
rewritten as follows \cite{GK0710}:
\bea
h\Phi  & = & \left(
\begin{array}{cccc}
-\phi' & 0 & 0& 0  \\
0  & 0_l & 0 & 0  \\
0 & 0 & h \Phi_n & 0 \\
0 & 0 & 0 & \frac{\mu^2}{\mu_{\phi}} 
{\bf 1}_{N_c-l-n} 
\end{array}
\right), \nonu \\
q  \widetilde{q} & = & \left(
\begin{array}{cccc}
\mu^2 {\bf 1}_{\widetilde{N}_c}+ \alpha \Lambda^2  \phi' & 0 & 0 & 0  \\
0 & \mu^2 {\bf 1}_l & 0  & 0 \\
0 & 0 & { \varphi}  \widetilde{\varphi}  &  0 \\
0 & 0 & 0 & 0_{N_c-l-n}
\end{array}
\right)
\label{Eigensec2}
\eea
with $\phi' = \mbox{diag} (\phi'_1, \cdots, \phi'_{\widetilde{N}_c})$.
Here $\varphi$ and $\widetilde{\varphi}$ are 
$n \times (\widetilde{N}_c-l)$
matrices and correspond to $n$-flavors of fundamentals of
the gauge group $SU(\widetilde{N}_c-l)$ which is unbroken.
One can move $n$ D4-branes from $(N_c-l)$ 
flavor D4-branes stretched
between the $D6_{-\theta}$-branes  and the NS5'-branes 
at $w=+v_{D6_{-\theta}}
\cot \theta $, to the local minimum of the potential and the end
points of these $n$ D4-branes are at a nonzero $w$ \cite{GK0710-1}.
In the brane configuration from Figure 2B, 
$\varphi$ and $\widetilde{\varphi}$ correspond to 
fundamental strings connecting between the $n$-flavor D4-branes and
$(\widetilde{N}_c-l)$-color D4-branes. Moreover,
the $h\Phi_n$ and ${ \varphi} 
\widetilde{\varphi}$
are $n \times n$ matrices.
The supersymmetric ground state corresponds to the vacuum expectation
values by
$h\Phi_n=\frac{\mu^2}{\mu_{\phi}}
{\bf 1}_{n}$ and $
\varphi \widetilde{\varphi} =0$. 

The full one loop potential from (\ref{Dualsec2}) and (\ref{Eigensec2}) 
takes the form 
\bea
\frac{V}{|h|^2}  & = &  
|\Phi_n  \varphi +\varphi \phi' |^2   
+  |\widetilde{\varphi} \Phi_n + \phi' \widetilde{\varphi}   |^2
  +  
\left| \varphi \widetilde{\varphi}-\mu^2 {\bf 1}_{n} + 
h 
\mu_{\phi}  \Phi_n \right|^2 
+ 
b |h \mu|^2 \tr \Phi_n^{\dagger}
\Phi_n 
\label{potsec2}
\eea
where $b =\frac{(\ln 4-1)}{8\pi^2}\widetilde{N}_c$
and we did not include $\Phi_n$- or $\Phi_n^{\dagger}$-independent term  
and after differentiating this (\ref{potsec2})
with respect to 
$\Phi_n^{\dagger}$
one obtains the local nonzero stable point 
given by
\bea
h \Phi_n 
\simeq \frac{\mu_{\phi}}{b}
{\bf 1}_n \qquad \mbox{or} \qquad
M_n \simeq \frac{\alpha  
\Lambda^3}{\widetilde{N}_c}  {\bf 1}_{n}.
\nonu
\eea
This corresponds to the location of $w$ coordinate, which is less
than $\frac{m}{\alpha \Lambda}(=\frac{\mu^2}{\mu_{\phi}})$, of 
 $n$ flavor D4-branes between 
the $D6_{-\theta}$-branes and the NS5'-branes.

$\bullet$ Non-coincident $N_f$ $D6_{-\theta}$-branes 
and $k'$ NS5'-branes(different massive
flavors) 

Let us suppose that the numbers of $D6_{-\theta}$-branes and 
the NS5'-branes are equal to each other:$N_f=k'$.
Let us displace the $k'$ $D6_{-\theta}$-branes and NS5'-branes
given in Figure 2A
in the $v$ direction respectively
to two $k'$ different points denoted by 
$v_{D6_{-\theta,j}}$ and $v_{NS5_{j}'}$
where $j=1,2, \cdots, k'$.  
The color and flavor D4-branes attached to them are displaced also.
The number of color D4-branes
stretched between $j$-th $NS5_j'$-brane and the NS5-brane
is denoted by $\widetilde{N}_{c,j}$
while the number of flavor D4-branes
stretched between the  $j$-th $D6_{-\theta,j}$-brane and  the 
$j$-th $NS5_{j}'$-brane
is denoted by $N_{f,j}$.
Then it is obvious that there are relations between the color and flavor
D4-branes $
\sum_{j=1}^{k'} \widetilde{N}_{c,j}
= \widetilde{N}_c$ and $
\sum_{j=1}^{k'} N_{f,j}= N_f$.
When all the $NS5_{j}'$-branes and 
$D6_{-\theta,j}$-branes 
are distinct, the low energy  
physics corresponds to $k'$-decoupled supersymmetric 
gauge theories with gauge group 
$
\prod_{j=1}^{k'} SU( \widetilde{N}_{c,j}) $. 

One deforms the Figure 2B by displacing the multiple
$D6_{-\theta}$-branes and $NS5'$-branes 
along $v$ direction, as in \cite{Ahn08-2}.
Then the $n$ curved flavor D4-branes attached to them(as well as other
D4-branes) are displaced
also as $k'$ different $n_j$'s connecting between 
$D6_{-\theta,j}$-brane and
$NS5_{j}'$-brane.
When we rescale the submeson field as
$M_{j} = h \Lambda \Phi_{j} $ \cite{Ahn08-2},
then the Kahler potential for $\Phi_{j}$ is canonical and the magnetic
quarks $q_j$ and $\widetilde{q}_j$ 
are canonical near the origin of field space \cite{ISS}.
Then the magnetic superpotential (\ref{Dualsec2}) 
can be rewritten in terms of $\Phi_j, q_j, \widetilde{q}_j$ and 
$ \phi_{\widetilde{N}_{c,j}}'$ which is 
coming from 4-4 strings connecting between the $j$-th 
$\widetilde{N}_{c,j}$ D4-branes
\bea
W_{dual} = \sum_{j=1}^{k'} \left[
 h \Phi_j  q_j   \widetilde{q}_j 
 +  
\frac{ \mu_{\phi}}{2} h^2 \tr 
\Phi_j^2- h \tr \mu_j^2  \Phi_j  + \frac{g_{\phi'}}{2} 
\tr {\phi^{'}}^{k'+1}_{\widetilde{N}_{c,j}}+  q_j
  \phi_{\widetilde{N}_{c,j}}' 
\widetilde{q}_j \right]
\label{dDual}
\eea
with
$
\mu_j^2 = m_j \Lambda_j$ and 
$\mu_{\phi} = \alpha \Lambda^2$ as before. 

One splits 
the $(N_{c,j}-l_j) \times (N_{c,j}-l_j)$
block  at the lower right corner of $h\Phi_j$ and $q_j
\widetilde{q}_j$ 
into blocks of 
size $n_j$ and $(N_{c,j}-l_j-n_j)$ for all $j$ 
as follows \cite{Ahn08-2}:
\bea
h \Phi & = &  \left(
\begin{array}{ccccc}
0_{\widetilde{N}_c+l+n} & 0 & 0 & \cdots  & 0 \\
0 & \frac{\mu_1^{2}}{\mu_{\phi}} 
{\bf 1}_{N_{c,1}- l_1-n_1} & 0 & \cdots & 0  \\
\cdot & \cdot & \cdot & \cdots & 0 \\
0 & 0 &  0 & 0 & \frac{\mu_{k'}^{2}}{\mu_\phi} 
{\bf 1}_{N_{c,k'}-l_{k'}-n_{k'}}  
\end{array}
\right) \nonu \\
&- &  \mbox{diag} 
(\phi_{\widetilde{N}_{c,1}}', 
\cdots, \phi_{\widetilde{N}_{c,k'}}', 0_{N_c}) 
 +  
\mbox{diag} (0_{\widetilde{N}_c+l}, h \Phi_{n_1}, 
\cdots, h \Phi_{n_{k'}}, 0_{N_c-l-n})
\label{M02}
\eea
and
\bea
q  \widetilde{q}   & = &  \left(
\begin{array}{ccccccc}
0_{\widetilde{N}_c} & 0 & 0 & 0 & 0 & 0 & 0 \\
0 & \mu_1^2  {\bf 1}_{l_1} & 0 & 0 & \cdots & 0  & 0  \\
0 & \cdot  & \cdot & \cdot & \cdots & \cdot  & 0 \\
0 & 0 & 0 & 0 & \cdots & \mu_{k'}^2  {\bf 1}_{l_{k'}}  & 0 \\ 
0 & 0 & 0 & 0 & 0 & 0 & 0_{N_c-l} 
\end{array}
\right) \nonu \\
& + &  
  \mbox{diag} 
(\mu_1^2 {\bf 1}_{\widetilde{N}_{c,1}} +\alpha \Lambda^2 
\phi_{\widetilde{N}_{c,1}}', 
\cdots, \mu_{k'}^2 {\bf 1}_{\widetilde{N}_{c,k'}} +\alpha \Lambda^2  
\phi_{\widetilde{N}_{c,k'}}', 0_{N_c}) \nonu \\
&+&
 \mbox{diag} (
0_{\widetilde{N}_c+l}, \varphi_{n_1} 
\widetilde{\varphi}_{n_1}, 
\cdots,  \varphi_{n_{k'}} \widetilde{\varphi}_{n_{k'}}, 
0_{N_c-l-n})
\label{solqq2}
\eea
where $l =\sum_{j=1}^{k'} l_j$ and $n =\sum_{j=1}^{k'} n_j$.
Here 
$\varphi_{n_j}$ and $\widetilde{\varphi}_{n_j}$ are 
$n_j \times (\widetilde{N}_{c,j}-l_j)$
matrices and correspond to $n_j$-flavors of fundamentals of
the gauge group $SU(\widetilde{N}_{c,j}-l_j)$ which is unbroken.
They correspond to 
fundamental strings connecting between the $n_j$-flavor D4-branes and
$(\widetilde{N}_{c,j}-l_j)$-color D4-branes. Moreover,
the $\Phi_{n_j}$ and ${ \varphi}_{n_j} 
\widetilde{\varphi}_{n_j}$
are $n_j \times n_j$ matrices.
The supersymmetric ground state corresponds to the vacuum expectation
values by
$h\Phi_{n_j}=\frac{\mu_j^2}{\mu_{\phi}}
{\bf 1}_{n_j}$ and $
\varphi_{n_j} \widetilde{\varphi}_{n_j}=0$.
The full one loop potential from (\ref{dDual}), (\ref{M02}) and 
(\ref{solqq2}) can be written similarly
and the local nonzero stable point arises as
\bea
h \Phi_{n_j} 
\simeq \frac{\mu_{\phi}}{b_j}
{\bf 1}_{n_j} \qquad \mbox{or} \qquad
M_{n_j} \simeq \frac{\alpha  
\Lambda^3}{\widetilde{N}_{c,j}}  {\bf 1}_{n_j}
\nonu
\eea
corresponding to the nonzero $w$ coordinates of $n_j$ flavor D4-branes between 
the $D6_{-\theta,j}$-brane and the $NS5_j'$-brane.

Therefore, the meta-stable states, for fixed $k'$ which is related to 
the order of the adjoint field in the superpotential  and $\theta$ which is a
deformation parameter by rotation angle of $D6_{-\theta}$-branes, 
are classified by the number of various D4-branes and the positions of
multiple $D6_{-\theta}$-branes and NS5'-branes: 
$
(N_{c,j}, N_{f,j}, l_j, n_j)$ and
$(v_{D6_{-\theta,j}}, v_{NS5_{j}'})$  where $j=1, 2, \cdots, k'$.

When $N_f$ is not equal to $k'$(for example, when $N_f > k'$), 
then some coincident $D6_{-\theta}$-branes among $N_f$ $D6_{-\theta}$-branes
should be connected to the NS5'-branes in order to connect all the
flavor D4-branes between the $D6_{-\theta}$-branes and the NS5'-branes.  
These  coincident $D6_{-\theta}$-branes can be obtained by taking those
quark masses equal. Then all the previous descriptions for the
meta-stable states for the case with $N_f=k'$ can be applied in this
case also without any difficulty.   
On the other hand, when $N_f < k'$, then 
some coincident NS5'-branes among $k'$ NS5'-branes
should be connected to the $D6_{-\theta}$-branes.

\section{$Sp(N_c)$ with two adj. and $N_f$-fund. }


\subsection{Electric theory}

The type IIA supersymmetric electric
brane configuration \cite{EGKRS,Ahn9712,AOT9712} corresponding to 
${\cal N}=1$ $Sp(N_c)$ gauge theory  with 
two adjoint fields $\Phi$, $\Phi'$(which are symmetric tensors in the
symplectic gauge group) and 
$N_f$-fundamental flavors $Q$
can be described as follows: one NS5-brane, $(2k'+1)$
NS5'-branes, 
$2N_c$
D4-branes, $2N_f$ D6-branes 
and the $O4^{\pm}$-planes(01236). 
The $2N_c$-color D4-branes are suspended between 
the left NS5-brane and the right NS5'-branes and the $2N_f$ D6-branes 
are located between the left NS5-brane and the right NS5'-branes.
The
$2N_f$-fields $Q$ are strings
stretching between $2N_f$ D6-branes and $2N_c$-color D4-branes. 
The adjoint field $\Phi'$
is related to the fluctuations of color D4-branes in $v$ direction. 

Let us deform this theory
by adding the mass 
and the quartic terms for fundamental quarks.
The former can be achieved by displacing the D6-branes along $ \pm v$
direction leading to their coordinates $v = \pm v_{D6}$ \cite{GK98} 
while the latter can be obtained by rotating the D6-branes
\cite{GK0710-1} 
by an angle 
$-\theta$ in $(w,v)$-plane and we denote them by $D6_{-\theta}$-branes. 
That is, each $N_f$ D6-branes is moving to the $ \pm v$
directions respectively and then is rotating 
by an angle $-\theta$ in $(w,v)$-plane.
Therefore, in the electric gauge theory, the deformed superpotential is
given by
\bea
W_{elec} = \left[ \frac{g_{\Phi'}}{2} \tr {\Phi'}^{2k'+2}+  Q \Phi'
  Q 
\right]  - m \tr Q Q +
\frac{\alpha}{2} \tr (Q Q)^2,
\qquad \alpha = \frac{\tan \theta}{\Lambda}, \quad 
m = \frac{v_{D6_{-\theta}}}{2\pi \ell_s^2}.
\label{superpotent}
\eea   
When $k'=0$, this theory reduces to the one similar to the unitary
case \cite{GK0710-1,GK0710}
because the first two terms of (\ref{superpotent}) contribute to
the additional quartic term for the quarks.
Therefore, we focus on the nontrivial case with $k' \geq 1$.
The order of the first term of (\ref{superpotent}) can be determined
by replacing the $k'$ of (\ref{electricsuperpotential}) with $(2k'+1)$
in the sense that when the additional $k'$ NS5'-branes are added into a
single NS5'-brane, their mirrors also should be present,
leading to the total number of NS5'-branes being $(2k'+1)$.

Let us summarize the ${\cal N}=1$ supersymmetric electric brane
configuration with superpotential 
(\ref{superpotent}) 
in type IIA string theory as follows and draw this in
Figure 3 which is nothing but the Figure 1 together with the addition 
of $O4^{\pm}$-planes:

$\bullet$ One left NS5-brane in (012345) directions with $w=0$

$\bullet$ $(2k'+1)$ right NS5'-branes in (012389) directions with $v=0$

$\bullet$ $2N_c$-color D4-branes in (01236) directions with $v=0=w$

$\bullet$ $2N_f$ $D6_{-\theta}$-branes in (01237)
directions and
two other directions in $(v,w)$-plane 

$\bullet$ $O4^{\pm}$-planes in (01236) directions with $v=0=w$ 

\begin{figure}[ht]
   \epsfxsize=1.5in 
\centerline{\epsffile{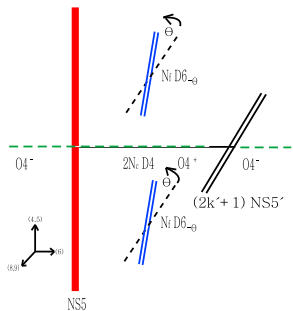}}
   \caption[FIG. \arabic{figure}.]{ 
The  ${\cal N}=1$ supersymmetric 
electric brane configuration for the gauge group $Sp(N_c)$ 
with two adjoints and fundamentals $Q$. 
Note that there are $(2k'+1)$ NS5'-branes and the fluctuations of 
$2N_c$ color D4-branes
in $v$ direction correspond to the adjoint field.
A 
rotation of $N_f$ D6-branes in $(w,v)$-plane
corresponds to 
a quartic term for the fundamentals while 
a displacement of $N_f$ D6-branes in $ \pm v$ direction corresponds to a
mass term for the fundamentals.
}
\end{figure}

\subsection{Magnetic theory}

Let us move the left NS5-brane to the right past the right 
NS5'-branes  as in previous section and we arrive at the Figure 4A.
Note that there exists a creation of $N_f$-flavor D4-branes
connecting $N_f$ $D6_{-\theta}$-branes and $(2k'+1)$ NS5'-branes(and their mirrors).
The linking number of the NS5-brane from Figure 4A
is 
$
l_m = \frac{2N_f}{2} -2 -2\widetilde{N}_c$.
The linking number of the NS5-brane from Figure 3
is
$
l_e = -\frac{2N_f}{2}+2 + 2N_c $.
From the equality between these, one obtains
the number of colors of dual magnetic theory \cite{EGKRS,ISS,Ahn06-1}
$
\widetilde{N}_c =N_f-N_c-2$.

\begin{figure}[ht]
   \epsfxsize=3.0in 
\centerline{\epsffile{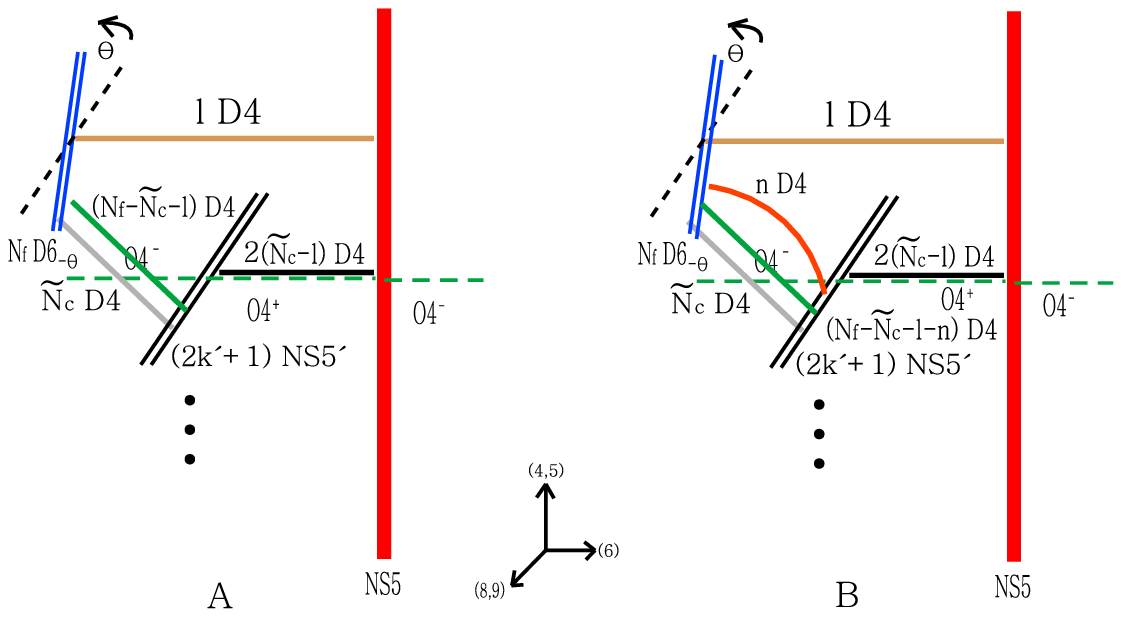}}
   \caption[FIG. \arabic{figure}.]{ 
The  ${\cal N}=1$ supersymmetric
magnetic brane configuration corresponding to Figure 3 with 
a misalignment between D4-branes when the gravitational potential of
the NS5-brane is ignored(4A) and nonsupersymmetric brane configuration
when  the gravitational potential of
the NS5-brane is considered(4B). 
The $N_f$ upper flavor D4-branes connecting between
upper $D6_{-\theta}$-branes and NS5'-branes are splitting into $ 
\widetilde{N}_c$-, $(N_f-\widetilde{N}_c-l)$- and
$l$- D4-branes(4A). Further $n$- D4-branes among 
upper $(N_f-\widetilde{N}_c-l)$- D4-branes are moved to the NS5-brane(4B). 
}
\end{figure}

The low energy theory on the color D4-branes 
has $Sp(\widetilde{N}_c)$ gauge group and an adjoint field 
$\phi'$ coming from 4-4 strings connecting the color D4-branes 
and 
$N_f$-fundamental quarks $q$
coming from 4-4 strings connecting between the color D4-branes and
flavor D4-branes.
Moreover, a single magnetic meson field $M \equiv Q Q$
is $2N_f \times 2N_f$ matrix and comes from 
4-4 strings of flavor D4-branes.
Then the magnetic superpotential is given by  
\bea
W_{mag} = \left[ \frac{g_{\phi'}}{2} \tr {\phi'}^{2k'+2}+  q \phi' q +
\frac{1}{\Lambda} M q q \right] 
+ \frac{\alpha}{2} \tr M^2 - m M.
\label{superpo1}
\eea 
When $k'=0$ and there is no mass term for $M$,
this superpotential reduces to the one found in 
\cite{ISS,Ahn06-1}.

In order to obtain the supersymmetric vacua, 
one computes the F-term equations for 
the superpotential (\ref{superpo1}):
\bea
 \frac{1}{\Lambda} M q + q \phi'& = & 0, \qquad
 g_{\phi'}(k'+1) {\phi'}^{2k'+1} + q q  =  0, \nonu \\
\qquad \frac{1}{\Lambda} q  q  & = &  m -\alpha M. 
\label{Fterm1}
\eea
We choose the adjoint field $\phi'$ to be diagonal, i.e., 
$\phi' = \mbox{diag} (\phi'_1, \cdots, \phi'_{\widetilde{N}_c})
\otimes \sigma_3$ \cite{Ahn9712}.
Then the first equation of (\ref{Fterm1}) implies that 
the upper left $2\widetilde{N}_c \times 2\widetilde{N}_c$ block 
of $\frac{M}{\Lambda}$ is given
by $ \mbox{diag} (-\phi'_1, \cdots, -\phi'_{\widetilde{N}_c}) \otimes
i \sigma_2$.
By substituting this into the last equation of (\ref{Fterm1}), one
obtains 
the upper left $2\widetilde{N}_c \times 2\widetilde{N}_c$ block 
of $\frac{1}{\Lambda} q q$ is given
by $ \mbox{diag} ( m+\alpha \Lambda \phi'_1, \cdots, 
m+ \alpha \Lambda \phi'_{\widetilde{N}_c}) \otimes i \sigma_2$.
Moreover, the lower right $2(N_f-\widetilde{N}_c) 
\times 2(N_f-\widetilde{N}_c)$ block 
of $M$ is given by $2l$'s zero eigenvalues and 
$2(N_f-\widetilde{N}_c-l)$'s eigenvalues
$\pm \frac{m}{\alpha}$.
Then finally, 
the second equation of (\ref{Fterm1}) gives rise to the expectation
value for the adjoint field $\phi'$ satisfying that 
\bea
 g_{\phi'}(k'+1) {\phi_j'}^{2k'+1}=-\Lambda(m+\alpha \Lambda \phi_j')
\label{condition1}
\eea
for $j=1,2, \cdots, \widetilde{N}_c$. 
For nonzero quark mass $m$, the expectation 
value $\phi_j'$ is not vanishing and $-\Lambda \phi_j' > \frac{m}{\alpha}$ for
the positiveness of left hand side of (\ref{condition1}) when
$g_{\phi'}$ and $\phi_j'$ are real.

Let us first describe the nonsupersymmetric meta-stable states
and supersymmetric ones 
when all the  $2N_f$ $D6_{-\theta}$-branes and $(2k'+1)$ NS5'-branes
are coincident with each other.

$\bullet$ Two coincident $N_f$ $D6_{-\theta}$-branes and coincident 
$(2k'+1)$ NS5'-branes 

One writes $2N_f \times 2N_f$ matrix $M$  with $2\widetilde{N}_c$
eigenvalues by the diagonal elements for 
$ \pm \varphi'$, $2l$'s eigenvalues by the zeros 
and $2(N_f-\widetilde{N}_c-l)$ eigenvalues
$\pm \frac{m}{\alpha}$ 
as follows:
\bea
M = \left(
\begin{array}{ccc}
-\Lambda \varphi' \otimes i \sigma_2 & 0 & 0  \\
0 & 0_{2l} & 0 \\
0 & 0 
& \frac{m}{\alpha} {\bf 1}_{N_f-\widetilde{N}_c-l} \otimes i \sigma_2
\end{array}
\right), \quad 
\varphi' = \mbox{diag} (\phi'_1, \cdots, \phi'_{\widetilde{N}_c}).
\label{M01}
\eea 
Therefore, in the brane configuration of Figure 4A, the $l$ of the
upper $N_f$ flavor D4-branes are connected with $l$ of $\widetilde{N}_c$ color
D4-branes
and the resulting D4-branes stretch from the upper $D6_{-\theta}$-branes to
the NS5-brane directly and the intersection point between the 
$l$ upper D4-branes and the NS5-brane is given by $(v,
w)=(+v_{D6_{-\theta}}, 0)$.
Now the $(N_f-\widetilde{N}_c-l)$ upper flavor D4-branes between 
the $D6_{-\theta}$-branes and 
the NS5'-branes are related to the corresponding half eigenvalues 
of $M$:$\frac{m}{\alpha} {\bf 1}_{N_f-\widetilde{N}_c-l}$.
The intersection point between the 
$(N_f-\widetilde{N}_c-l)$ upper D4-branes and the NS5'-branes is given 
by $(v, w)=(0, +v_{D6_{-\theta}} \cot \theta)$ corresponding to
positive eigenvalues of $M$.
Finally, the remnant $2\widetilde{N}_c$-flavor D4-branes
between 
the $D6_{-\theta}$-branes and 
the NS5'-branes 
correspond to the eigenvalues $ \pm \varphi'$ in (\ref{M01}) with 
(\ref{condition1}) which shows the exact coordinate of $w$ for these
flavor D4-branes. 

One represents the vacuum expectation value for the quarks
as follows:
\bea
q  q = \left(
\begin{array}{ccc}
\Lambda (m {\bf 1}_{\widetilde{N}_c} 
+ \alpha \Lambda \varphi') \otimes i \sigma_2   & 0 & 0  \\
0 & \Lambda m {\bf 1}_{l} \otimes i \sigma_2 & 0 \\
0 & 0 & 0_{2(N_f-\widetilde{N}_c-l)} 
\end{array}
\right),  
\varphi' = \mbox{diag} (\phi'_1, \cdots, \phi'_{\widetilde{N}_c}).
\label{qq}
\eea
In the $l$-th vacuum the gauge symmetry is broken to
$Sp(\widetilde{N}_c-l)$ and when the supersymmetric vacuum is drawn in
Figure 4A with $l=0$, the gauge group 
$Sp(\widetilde{N}_c)$ is unbroken. 
Then the supersymmetric 
ground state corresponds to the vacuum expectation values $M$ by
$\pm \varphi'$ which has $2\widetilde{N}_c$'s eigenvalues and  
$\pm \frac{m}{\alpha}$ with degeneracy $2(N_f-\widetilde{N}_c)$.

Let us rescale the meson field as
$M = h \Lambda \Phi $ \cite{ISS}.
Then the magnetic superpotential (\ref{superpo1}) 
can be rewritten in terms of $\Phi, q$ and $\phi'$
\bea
W_{dual} = 
 h \Phi  q   q 
 +  
\frac{ \mu_{\phi}}{2} h^2 \tr \Phi^2- h \mu^2 \tr \Phi +
\frac{g_{\phi'}}{2} \tr {\phi'}^{2k'+2}+  q \phi' q
\label{Dual1}
\eea
where
$
\mu^2 = m \Lambda$ and  
$\mu_{\phi} = \alpha \Lambda^2$.

Now one splits 
the $2(N_f-\widetilde{N}_c-l) \times 2(N_f-\widetilde{N}_c-l)$
block  at the lower right corner of $M$ and $q q$ 
into blocks of 
size $2n$ and $2(N_f-\widetilde{N}_c-l-n)$ and then 
(\ref{M01}) and (\ref{qq}) are 
rewritten as follows:
\bea
h\Phi & = & \left(
\begin{array}{cccc}
-\varphi' \otimes i \sigma_2 & 0 & 0 & 0  \\
0 & 0_{2l} & 0 & 0 \\
0 & 0 & h \Phi_{2n} & 0 \\
0 & 0 & 0 & \frac{\mu^2}{\mu_{\phi}} 
{\bf 1}_{(N_f-\widetilde{N}_c-l-n)} \otimes i \sigma_2
\end{array}
\right), \nonu \\
q  q & = & \left(
\begin{array}{cccc}
(\mu^2 {\bf 1}_{\widetilde{N}_c} + \alpha \Lambda^2  \varphi') 
\otimes i \sigma_2 & 0 & 0 & 0  \\
0 & \mu^2 {\bf 1}_{l} \otimes i \sigma_2 & 0 & 0 \\
0 & 0 & { \varphi}  \varphi  &  0 \\
0 & 0 & 0 & 0_{2(N_f-\widetilde{N}_c-l-n)}
\end{array}
\right).
\nonu
\eea
Here $\varphi$  is $2n \times 2(\widetilde{N}_c-l)$
dimensional matrix and corresponds to $2n$ flavors of fundamentals of
the gauge group $Sp(\widetilde{N}_c-l)$ which is unbroken by the nonzero
expectation value of $q$.
The $\Phi_{2n}$ and $ \varphi \varphi$
are $2n \times 2n$ matrices.
The supersymmetric ground state corresponds to
the vacuum expectation values by
$h\Phi_{2n}= \frac{\mu^2}{\mu_{\phi}} {\bf 1}_{n} \otimes i \sigma_2$
and $\varphi =0$. 

The full one loop potential takes the form 
\bea
\frac{V}{|h|^2}  & = &  
|\Phi_{2n}  \varphi +\varphi \varphi' |^2   
  +  
\left| \varphi \varphi-\mu^2 {\bf 1}_{2n} + 
h 
\mu_{\phi}  \Phi_{2n} \right|^2 
+ b |h \mu|^2 \tr \Phi_{2n}
\Phi_{2n} 
\nonu
\eea
where $b =\frac{(\ln 4-1)}{8\pi^2}\widetilde{N}_c$ 
and we did not include $\Phi_{2n}$-independent term 
and after differentiating this 
with respect to 
$\Phi_{2n}$
one obtains the local nonzero stable point 
given by
\bea
h \Phi_{2n} 
\simeq \frac{\mu_{\phi}}{b}
{\bf 1}_n \otimes i \sigma_2 \qquad \mbox{or} \qquad
M_{2n} \simeq \frac{\alpha  
\Lambda^3}{\widetilde{N}_c}  {\bf 1}_{n} \otimes i \sigma_2
\nonu
\eea
corresponding to the location of $w$ coordinates for
$n$ flavor D4-branes, very close to the NS5-brane, between 
the $D6_{-\theta}$-branes and the NS5'-branes.

$\bullet$ Non-coincident $D6_{-\theta}$-branes and NS5'-branes 

Let us consider when the numbers of $D6_{-\theta}$-branes and
the number of NS5'-branes minus one are equal:$N_f=k'$.
We displace the $k'$ upper $D6_{-\theta}$-branes and upper NS5'-branes, 
given in Figure 4A, in the $+v$ direction respectively
to two $k'$ different points denoted by 
$v_{D6_{-\theta,j}}$ and $v_{NS5_{j}'}$
where $j=1,2, \cdots, k'$(and their mirrors in the $-v$ direction).
A single  NS5'-brane is located at $v=0$.
Then there are 
$
\sum_{j=1}^{k'} \widetilde{N}_{c,j}(\equiv N_{f,j}-N_{c,j}-2)
= \widetilde{N}_c$ and $
\sum_{j=1}^{k'} N_{f,j} = N_f$.
When all the upper $NS5_{j}'$-branes and 
$D6_{-\theta,j}$-branes($j=1, 2, \cdots, k'$) 
are distinct, the low energy  
physics corresponds to $k'$ decoupled supersymmetric 
gauge theories with gauge groups 
$
\prod_{j=1}^{k'} Sp( \widetilde{N}_{c,j})$.

One deforms the Figure 4B by displacing the multiple
$D6_{-\theta}$-branes and $NS5'$-branes 
along $v$ direction \cite{Ahn08-2}.
Then the $n$ curved flavor D4-branes attached to them as well as other
D4-branes are displaced
also as $k'$ different $n_j$'s connecting between 
$D6_{-\theta,j}$-brane and
$NS5_{j}'$-brane.
Then the magnetic superpotential (\ref{Dual1}) 
can be rewritten in terms of $\Phi_j, q_j$ and 
$ \phi_{\widetilde{N}_{c,j}}'$ which is 
coming from 4-4 strings connecting between the $j$-th 
$\widetilde{N}_{c,j}$ D4-branes
\bea
W_{dual} = \sum_{j=1}^{k'} \left[
 h \Phi_j  q_j  q_j 
 +  
\frac{ \mu_{\phi}}{2} h^2 \tr 
\Phi_j^2- h \tr \mu_j^2  \Phi_j  + \frac{g_{\phi'}}{2} 
\tr {\phi^{'}}^{2k'+2}_{\widetilde{N}_{c,j}}+  q_j
  \phi_{\widetilde{N}_{c,j}}' 
q_j \right]
\label{dDualnew}
\eea
with
$
\mu_j^2 = m_j \Lambda_j$ and 
$\mu_{\phi} = \alpha \Lambda^2$ as before. 

One splits 
the $2(N_{f,j}-\widetilde{N}_{c,j}-l_j) \times 2(N_{f,j}-
\widetilde{N}_{c,j}-l_j)$
block  at the lower right corner of $h\Phi_j$ and $q_j
q_j$ 
into blocks of 
size $2n_j$ and $2(N_{f,j}-\widetilde{N}_{c,j}-l_j-n_j)$ for all $j$ 
as follows:
\bea
h \Phi & = &  \left(
\begin{array}{ccccc}
0_{2(\widetilde{N}_c+l+n)} & 0 & 0 & \cdots  & 0 \\
0 & \frac{\mu_1^{2}}{\mu_{\phi}} 
{\bf 1}_{N_{f,1} - \widetilde{N}_{c,1}- l_1-n_1} \otimes i \sigma_2 & 0 & \cdots & 0  \\
\cdot & \cdot & \cdot & \cdots & 0 \\
0 & 0 &  0 & 0 & \frac{\mu_{k'}^{2}}{\mu_\phi} 
{\bf 1}_{N_{f,k'}-\widetilde{N}_{c,k'}-l_{k'}-n_{k'}} \otimes i \sigma_2 
\end{array}
\right) \nonu \\
&- &  \mbox{diag} 
(\phi_{\widetilde{N}_{c,1}}', 
\cdots, \phi_{\widetilde{N}_{c,k'}}', 0_{N_f-\widetilde{N}_c}) \otimes i \sigma_2 
 \nonu \\
&+&  
\mbox{diag} (0_{2(\widetilde{N}_c+l)}, h \Phi_{2n_1}, 
\cdots, h \Phi_{2n_{k'}}, 0_{2(N_f-\widetilde{N}_c-l-n)})
\label{M02new}
\eea
and
\bea
q  q   & = &  \left(
\begin{array}{ccccccc}
0_{2\widetilde{N}_c} & 0 & 0 & 0 & 0 & 0 & 0 \\
0 & \mu_1^2  {\bf 1}_{2l_1} & 0 & 0 & \cdots & 0  & 0  \\
0 & \cdot  & \cdot & \cdot & \cdots & \cdot  & 0 \\
0 & 0 & 0 & 0 & \cdots & \mu_{k'}^2  {\bf 1}_{2l_{k'}}  & 0 \\ 
0 & 0 & 0 & 0 & 0 & 0 & 0_{2(N_f-\widetilde{N}_c-l)} 
\end{array}
\right) \nonu \\
& + &  
  \mbox{diag} 
(\mu_1^2 {\bf 1}_{\widetilde{N}_{c,1}} + \alpha \Lambda^2 
\phi_{\widetilde{N}_{c,1}}', 
\cdots, \mu_{k'}^2 {\bf 1}_{\widetilde{N}_{c,k'}} + \alpha \Lambda^2 
\phi_{\widetilde{N}_{c,k'}}', 0_{N_f-\widetilde{N}_c}) \otimes i \sigma_2 \nonu \\
&+&
 \mbox{diag} (
0_{2(\widetilde{N}_c+l)}, \varphi_{2n_1} 
\varphi_{2n_1}, 
\cdots,  \varphi_{2n_{k'}} \varphi_{2n_{k'}}, 
0_{2(N_f-\widetilde{N}_c-l-n)})
\label{solqq2new}
\eea
where $l =\sum_{j=1}^{k'} l_j$ and $n =\sum_{j=1}^{k'} n_j$.
Here 
$\varphi_{2n_j}$  is 
$2n_j \times 2(\widetilde{N}_{c,j}-l_j)$
matrices and correspond to $2n_j$-flavors of fundamentals of
the gauge group $Sp(\widetilde{N}_{c,j}-l_j)$ which is unbroken.
The supersymmetric ground state corresponds to the vacuum expectation
values by
$h\Phi_{2n_j}=\frac{\mu_j^2}{\mu_{\phi}}
{\bf 1}_{n_j} \otimes i \sigma_2$ and $
\varphi_{2n_j}=0$.
The full one loop potential from (\ref{dDualnew}), (\ref{M02new}) and 
(\ref{solqq2new}) can be written similarly
and the local nonzero stable point arises as
\bea
h \Phi_{2n_j} 
\simeq \frac{\mu_{\phi}}{b_j}
{\bf 1}_{n_j} \otimes i \sigma_2 \qquad \mbox{or} \qquad
M_{2n_j} \simeq \frac{\alpha  
\Lambda^3}{\widetilde{N}_{c,j}}  {\bf 1}_{n_j} \otimes i \sigma_2
\nonu
\eea
corresponding to the $w$ coordinates of $n_j$ flavor D4-branes between 
the $D6_{-\theta,j}$-brane and the $NS5_j'$-brane(and their mirrors).

Therefore, the meta-stable states, for fixed $k'$ and $\theta$, 
are classified by the number of various D4-branes and the positions of
multiple $D6_{-\theta}$-branes and NS5'-branes: 
When $N_f$ is not equal to $k'$(for example, when $N_f > k'$), 
then some coincident $D6_{-\theta}$-branes among $N_f$ $D6_{-\theta}$-branes
should be connected to the NS5'-branes in order to connect all the
flavor D4-branes between the $D6_{-\theta}$-branes and the NS5'-branes.  

%

\section{$SU(N_c)$ with $N_f$-fund. and a symm.}


\subsection{Electric theory}

The type IIA supersymmetric electric
brane configuration \cite{LLL,Ahn07,GK98} corresponding to 
${\cal N}=1$ $SU(N_c)$ gauge theory  with a symmetric tensor $S$,
a conjugate symmetric tensor $\widetilde{S}$, 
$N_f$-fundamental flavors $Q, \widetilde{Q}$
can be described as follows: one middle NS5-brane, $2k'$
NS5'-branes, 
$N_c$-color
D4-branes, $2N_f$ D6-branes 
and the $O6^{+}$-plane(0123789). 
The $N_c$-color D4-branes are suspended between 
the left $k'$ NS5'-branes  and the right $k'$ NS5'-branes 
and similarly each $N_f$ D6-branes 
is located between the middle NS5-brane and the NS5'-branes.

Let us deform this theory \cite{ILS}
by adding the mass  
and the quartic terms for fundamental quarks.
The former can be achieved by displacing the D6-branes along $ \pm v$
direction leading to their coordinates $v = \pm v_{D6}$ \cite{GK98} 
while the latter can be obtained by rotating the D6-branes
\cite{GK0710-1} 
by the angles 
$ \mp \theta$ in $(w,v)$-plane and we denote them by $D6_{ \mp \theta}$-branes. 
Therefore, in the electric gauge theory, the deformed superpotential is
given by
\bea
W_{elec} = \frac{\alpha}{2} \tr (Q \widetilde{Q})^2 - m \tr Q
\widetilde{Q} +\left[-\frac{\beta}{2}   \tr (S \widetilde{S})^{k'+1} + m_S
\tr S \widetilde{S}\right].
\label{potential}
\eea 
Here the $\alpha=\frac{\tan \theta}{\Lambda}$ and 
$m=\frac{v_{D6_{-\theta}}}{2\pi \ell_s^2}$ 
correspond to  the rotation and displacement 
of D6-branes and they are
the same as the ones in \cite{Ahn07-11}.
The last two terms in (\ref{potential}) 
are due to the rotation of NS5'-branes where
$\beta=\tan \omega$ and $\omega$ is a rotation angle of NS5'-branes 
in the $(w,v)$-plane
and the relative displacement of D4-branes where $m_S =v_{NS5'}$ is
the distance of D4-branes in $v$ direction.
When $k'=1$ with small $\beta$ limit 
and $m_S=0$, this theory reduces to the one \cite{Ahn07-11}.
Therefore, we focus on the case with $k' \geq 2$ and 
$\beta, m_S \rightarrow  0$, i.e., we consider the multiple
NS5'-branes with the electric 
superpotential (\ref{potential}) with this particular limit.

Let us summarize the ${\cal N}=1$ supersymmetric electric brane
configuration with superpotential 
(\ref{potential}) 
in type IIA string theory as follows and draw this in
Figure 5:

$\bullet$ One NS5-brane in (012345) directions with $w=0=x^6$

$\bullet$ $2k'$ NS5'-branes in (012389) directions with $v=0$

$\bullet$ $N_c$-color D4-branes in (01236) directions with $v=0=w$

$\bullet$ $N_f$ $D6_{\pm \theta}$-branes in (01237)
directions and
two other directions in $(v,w)$-plane

$\bullet$ $O6^{+}$-plane in (0123789) directions with $x^6=0=v$ 

\begin{figure}[ht]
   \epsfxsize=2.0in 
\centerline{\epsffile{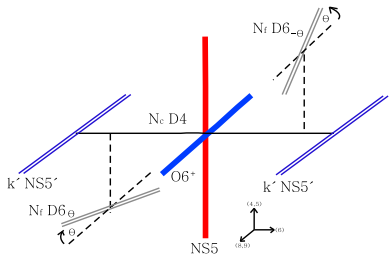}}
   \caption[FIG. \arabic{figure}.]{ 
The  ${\cal N}=1$ supersymmetric 
electric brane configuration for the gauge group $SU(N_c)$ 
with a symmetric tensor and fundamentals $Q, \widetilde{Q}$. 
Note that there are multiple $k'$ NS5'-branes.
A 
rotation of $N_f$ D6-branes in $(w,v)$-plane, denoted by $D6_{\mp \theta}$-branes,
corresponds to 
a quartic term for the fundamentals while 
a displacement of $N_f$ D6-branes in $ \pm v$ direction corresponds to a
mass term for the fundamentals.
}
\end{figure}

\subsection{Magnetic theory}

We move the $D6_{\mp \theta}$-branes and
the left and right 
NS5'-branes through each other and use the linking numbers for the
computation of creation of D4-branes as in \cite{Ahn07}.
Let us take the ``extra'' left- and right-$(k'-1)$ NS5'-branes,
compared with the single NS5'-brane case \cite{Ahn07,Ahn07-11}, move
them to the origin
 $x^6=0$, and rotate them by an angle $\frac{\pi}{2}$ in $(w,v)$-plane, 
coinciding with the middle NS5-brane. Temporarily, there
exist $(2k'-1)$ middle NS5-branes at $x^6=0$.
Let us move the left $D6_{\theta}$-branes to the right all the way(and their
mirrors, right $D6_{-\theta}$-branes to the left) past the NS5-branes
and the right single NS5'-brane. 
Then the linking number $l_m$ of a $D6_{\theta}$-brane
becomes $l_m=\frac{1}{2}-n_{4L}$ while 
the one of the same $D6_{\theta}$-brane $l_e=-\frac{1}{2}$ in the
electric theory.  
Then the number of D4-branes to the left of this $D6_{\theta}$-brane 
$n_{4L}$ in the magnetic theory becomes 1 and we must add $N_f$ D4-branes to the
left side of all $N_f$ $D6_{\theta}$-branes(and their mirrors).  
Note that at the $x^6=0$, the $D6_{\pm \theta}$-branes become the
$D6_{\pm \frac{\pi}{2}}$-branes instantaneously \cite{Ahn07,Ahn07-11}. 

Then the extra $2(k'-1)$ middle NS5-branes, which were 
left- and right-$(k'-1)$ NS5'-branes in the electric theory,
are moving to $\pm x^6$ direction by performing the remaining dual
process and rotating 
by an angle $\frac{\pi}{2}$ in $(w,v)$-plane. This 
leads to the left $k'$ NS5'-branes  and the right $k'$ 
NS5'-branes which look similar to
the brane configuration of electric theory but we need to further take Seiberg
dual for the remaining single NS5'-brane here.  
In this process there is a creation of D4-branes because 
the O6-plane, which has 4 D6-brane charge, is not parallel to 
these $2(k'-1)$ NS5-branes at $x^6=0$. 
Finally, after moving the left NS5'-brane, which does not participate in the
dual process so far, to the right all the way past
O6-plane(and its mirror, the right NS5'-brane to the left), 
the linking number
of the NS5'-brane can be computed as  
$l_m =\frac{N_f}{2k'}+
\frac{4(k'-1)}{2k'}-\frac{\widetilde{N}_c}{k'}+  \frac{N_f}{k'}$.
Note that the fractional contribution $\frac{k'-1}{k'}$
in the second term is due to the
fact that only the left $(k'-1)$ NS5'-branes among $k'$ NS5'-branes 
in the electric theory are crossing 
the O6-plane with an angle.
Of course,  the remaining left single NS5'-brane among $k'$ NS5'-branes 
in the electric theory is crossing 
the $N_f$ $D6_{-\theta}$-branes with an angle.
Originally, the linking number was given by 
$l_e=-\frac{N_f}{2k'}-\frac{4(k'-1)}{2k'}+\frac{N_c}{k'}$.
This leads to the fact that the number of D4-branes becomes
$
\widetilde{N}_c = 2N_f-N_c+4(k'-1)$.
Here the last constant term which depends on the number of NS5'-branes
is due to the fact that 
the ``extra'' NS5'-branes are crossing the O6-plane.
Of course, $k'=1$ case reduces to the known number of D4-branes 
$\widetilde{N}_c = 2N_f -N_c $ \cite{Ahn07}. 
We present the Figure 6A.

\begin{figure}[ht]
   \epsfxsize=3.0in 
\centerline{\epsffile{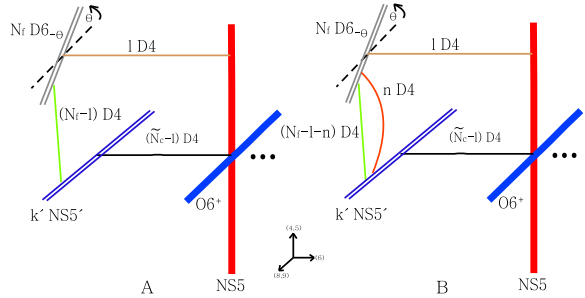}}
   \caption[FIG. \arabic{figure}.]{ 
The  ${\cal N}=1$ supersymmetric
magnetic brane configuration corresponding to Figure 5 with 
a misalignment between D4-branes when the gravitational potential of
the NS5-brane is ignored(6A) and nonsupersymmetric brane configuration
when  the gravitational potential of
the NS5-brane is considered(6B). 
The $N_f$ flavor D4-branes connecting between
$D6_{-\theta}$-branes and NS5'-branes are splitting into $(N_f-l)$- and
$l$- D4-branes(6A). Further $n$- D4-branes among 
$(N_f-l)$- D4-branes are moved to the NS5-brane(6B). 
}
\end{figure}

The low energy theory on the color D4-branes 
has $SU(\widetilde{N}_c)$ gauge group and  
$N_f$-fundamental dual quarks $q, \widetilde{q}$
coming from 4-4 strings connecting between the color D4-branes and
flavor D4-branes and a symmetric tensor 
and a conjugate symmetric tensor  $s, \widetilde{s}$
coming from 4-4 strings connecting between the color D4-branes with
$x^6 < 0$
and
the color D4-branes with $x^6 > 0$.
Moreover, a single magnetic meson field $M \equiv Q \widetilde{Q}$
is $N_f \times N_f$ matrix and comes from 
4-4 strings of flavor D4-branes.
Then the magnetic superpotential with the limit $\beta, m_S
\rightarrow 0$ 
is given by  
\bea
W_{mag} = \frac{1}{\Lambda} M q  (\widetilde{s} s)^{k'} \widetilde{q} +  
\frac{\alpha}{2} \tr M^2- m \tr M. 
\label{magsymm}
\eea
The case where $k'=1$ and $\alpha=0$ leads to the previous result in \cite{Ahn07}.
Note that the number of color $\widetilde{N}_c$ depends on $k'$.
In general, there are also different kinds of meson fields 
$M_j = Q (\widetilde{S} S)^j \widetilde{Q}, P_r = Q (\widetilde{S}
S)^r \widetilde{S} Q $ and $ \widetilde{P}_r = \widetilde{Q} S (\widetilde{S}
S)^r  \widetilde{Q}$ where $j=1, 2, \cdots,
k'$ and $r=0, 1, \cdots, k'-1$ for general rotation angles of
multiple NS5'-branes.
The magnetic superpotential contains the interaction between these
meson fields with $q, \widetilde{q}, s$ and $\widetilde{s}$ \cite{ILS}
as well as $(s \widetilde{s})^{k'+1}$ which will vanish for small
$\beta$ limit. 
However, the particular route we take above from an electric theory to
the magnetic theory does not produce these meson fields because the
$N_f$ $D6_{\pm \theta}$-branes and the extra $2(k'-1)$ NS5'-branes do not
create the D4-branes corresponding to those meson fields.  

For the supersymmetric vacua, one can compute the F-term equations for
this superpotential (\ref{magsymm}) 
and the expectation values for $M$ and $q
\widetilde{s} s \widetilde{q}$ are obtained since this superpotential
is the same as the one in \cite{Ahn07-11}.
Let us consider two different cases.

$\bullet$ Coincident $N_f$ $D6_{-\theta}$-branes and $k'$ 
NS5'-branes(and their mirrors) 

In this case, all the discussions given in \cite{Ahn07-11}
are satisfied with one exception that the number of colors 
here is different from the one in \cite{Ahn07-11}. 
The theory has many nonsupersymmetric meta-stable ground states and 
when we rescale the meson field as
$
M = h \Lambda \Phi $,
then the Kahler potential for $\Phi$ is canonical and the magnetic
quarks are canonical near the origin of field space \cite{ISS}.
Then the magnetic superpotential can be written in terms of $\Phi$ or $M$
\bea
W_{mag} = h \Phi  q  (\widetilde{s} s)^{k'} \widetilde{q} 
 +  
\frac{h^2 \mu_{\phi}}{2} \tr \Phi^2- h \mu^2 \tr \Phi.
\nonu
\eea
Now one splits 
the $(N_f-l) \times (N_f-l)$
block  at the lower right corner of $h\Phi$ and $q \widetilde{s} s
\widetilde{q}$ 
into blocks of 
size $n$ and $(N_f-l-n)$ as follows \cite{Ahn07-11}:
\bea
h\Phi = \left(
\begin{array}{ccc}
0_l & 0 & 0  \\
0 & h \Phi_n & 0 \\
0 & 0 & \frac{\mu^2}{\mu_{\phi}} {\bf 1}_{N_f-l-n}
\end{array}
\right), \qquad
q (\widetilde{s} s)^{k'} \widetilde{q} = \left(
\begin{array}{ccc}
\mu^2 {\bf 1}_l & 0 & 0  \\
0 & { \varphi} \widetilde{\beta} \beta \widetilde{\varphi}  &  0 \\
0 & 0 & 0_{N_f-l-n}
\end{array}
\right).
\nonu
\eea
Here $\varphi$ and $\widetilde{\varphi}$ are $n \times (\widetilde{N}_c-l)$
dimensional matrices and correspond to $n$ flavors of fundamentals of
the gauge group $SU(\widetilde{N}_c-l)$ which is unbroken.
In the brane configuration shown in Figure 6B, 
they correspond to 
fundamental strings connecting the $n$ flavor D4-branes and
$(\widetilde{N}_c-l)$
color D4-branes \cite{Ahn07-11}.
The $\Phi_n$ and ${ \varphi} \widetilde{\beta} \beta
\widetilde{\varphi}$
are $n \times n$ matrices.
The supersymmetric ground state corresponds to
$h\Phi_n= \frac{\mu^2}{\mu_{\phi}} {\bf 1}_{n}$ and $ 
\varphi \widetilde{\beta}=0=\beta \widetilde{\varphi}$. 

Now the full one loop potential 
for $\Phi_n, \hat{\varphi} \equiv \varphi \widetilde{\beta}$ and $ 
\hat{\widetilde{\varphi}} \equiv \beta \widetilde{\varphi} $
\cite{Ahn07} 
takes the form
\bea
\frac{V}{|h|^2}  =  
|\Phi_n  \hat{\varphi}|^2   
+  |\Phi_n  \hat{\widetilde{\varphi}}  |^2
  +  
| \hat{\varphi}  \hat{\widetilde{\varphi}}-\mu^2 {\bf 1}_{n} + 
h \mu_{\phi} \Phi_n|^2 + b |h \mu|^2 \tr \Phi_n^{\dagger} \Phi_n, 
\nonu
\eea
where $b = \frac{(\ln 4-1)}{8\pi^2} \widetilde{N}_c$.
Differentiating this potential with respect to 
$\Phi_n^{\dagger}$ and putting $\hat{\varphi}=0=\hat{
\widetilde{\varphi}}$, one obtains
\bea
h \Phi_n 
\simeq \frac{ \mu_{\phi}}{b }
{\bf 1}_n \qquad \mbox{or} \qquad
M_n \simeq \frac{\alpha \Lambda^3}{\widetilde{N}_c} {\bf 1}_{n}
\label{sec4vac}
\eea
corresponding to the $w$ coordinates of $n$ curved flavor D4-branes between 
the $D6_{-\theta}$-branes and the NS5'-branes(and their mirrors).

$\bullet$ Non-coincident $D6_{-\theta}$-branes and NS5'-branes 

Let us consider the case where the numbers of $D6_{-\theta}$-branes and 
the NS5'-branes are equal to each other:$N_f=k'$.
We displace the $k'$ $D6_{-\theta}$-branes and NS5'-branes
given in Figure 6A
in the $v$ direction respectively
to two $k'$ different points denoted by 
$v_{D6_{-\theta,j}}$ and $v_{NS5_{j}'}$
where $j=1,2, \cdots, k'$, as in section 2.  
There are relations between the color and flavor
D4-branes as follows: $
\sum_{j=1}^{k'} \widetilde{N}_{c,j}
= \widetilde{N}_c$ and $
\sum_{j=1}^{k'} N_{f,j}= N_f$.

One deforms the Figure 6B by displacing the multiple
$D6_{-\theta}$-branes and $NS5'$-branes 
along $v$ direction.
Then the $n$ curved flavor D4-branes attached to them are displaced
also as $k'$ different $n_j$'s connecting between 
$D6_{-\theta,j}$-brane and
$NS5_{j}'$-brane.
Let us rescale the submeson field as
$M_{j} = h \Lambda \Phi_{j} $ \cite{Ahn08-2} and 
the Kahler potential for $\Phi_{j}$ is canonical and the magnetic
quarks $q_j$ and $\widetilde{q}_j$ as well as $s_j$ and $\widetilde{s}_j$
are canonical near the origin of field space \cite{ISS}.
Then the magnetic superpotential 
can be rewritten in terms of $\Phi_j, q_j, \widetilde{q}_j, s_j$ and 
$ \widetilde{s}_j$ 
\bea
W_{mag} = \sum_{j=1}^{k'} 
\left[ h \Phi_j  q_j  (\widetilde{s}_j s_j)^{k'} \widetilde{q}_j 
 +  
\frac{h^2 \mu_{\phi}}{2} \tr \Phi_j^2- h \mu_j^2 \tr \Phi_j \right].
\nonu
\eea
with
$
\mu_j^2 = m_j \Lambda_j$ and 
$\mu_{\phi} = \alpha \Lambda^2$ as before. 

One splits 
the $(N_{f,j}-l_j) \times (N_{f,j}-l_j)$
block  at the lower right corner of $h\Phi_j$ and $q_j \widetilde{s}_j s_j
\widetilde{q}_j$ 
into blocks of 
size $n_j$ and $(N_{f,j}-l_j-n_j)$ for all $j$ 
as follows \cite{Ahn08-2}:
\bea
h \Phi & = &  \left(
\begin{array}{ccccc}
0_{l+n} & 0 & 0 & \cdots  & 0 \\
0 & \frac{\mu_1^{2}}{\mu_{\phi}} 
{\bf 1}_{N_{f,1}- l_1-n_1} & 0 & \cdots & 0  \\
\cdot & \cdot & \cdot & \cdots & 0 \\
0 & 0 &  0 & 0 & \frac{\mu_{k'}^{2}}{\mu_\phi} 
{\bf 1}_{N_{f,k'}-l_{k'}-n_{k'}}  
\end{array}
\right) \nonu \\
&+ &  
\mbox{diag} (0_l, h \Phi_{n_1}, 
\cdots, h \Phi_{n_{k'}}, 0_{N_f-l-n})
\nonu
\eea
and
\bea
q  (\widetilde{s} s)^{k'} \widetilde{q}   & = &  \left(
\begin{array}{cccccc}
 \mu_1^2  {\bf 1}_{l_1} & 0 & 0 & \cdots & 0  & 0  \\
 \cdot  & \cdot & \cdot & \cdots & \cdot  & 0 \\
 0 & 0 & 0 & \cdots & \mu_{k'}^2  {\bf 1}_{l_{k'}}  & 0 \\ 
 0 & 0 & 0 & 0 & 0 & 0_{N_f-l} 
\end{array}
\right) \nonu \\
&+&
 \mbox{diag} (
0_l, \varphi_{n_1} \widetilde{\beta}_{n_1} \beta_{n_1}
\widetilde{\varphi}_{n_1}, 
\cdots,  \varphi_{n_{k'}} \widetilde{\beta}_{n_k'} \beta_{n_k'} 
\widetilde{\varphi}_{n_{k'}}, 
0_{N_f-l-n})
\nonu
\eea
where $l =\sum_{j=1}^{k'} l_j$ and $n =\sum_{j=1}^{k'} n_j$, as before.
Here 
$\varphi_{n_j}$ and $\widetilde{\varphi}_{n_j}$ are 
$n_j \times (\widetilde{N}_{c,j}-l_j)$
matrices and correspond to $n_j$-flavors of fundamentals of
the gauge group $SU(\widetilde{N}_{c,j}-l_j)$ which is unbroken.
The supersymmetric ground state corresponds to the vacuum expectation
values by
$h\Phi_{n_j}=\frac{\mu_j^2}{\mu_{\phi}}
{\bf 1}_{n_j}$ and $
\varphi_{n_j} \widetilde{\beta}_{n_j}= \beta_{n_j} 
\widetilde{\varphi}_{n_j}=0$.
The full one loop potential can be written similarly
and the local nonzero stable point arises as
\bea
h \Phi_{n_j} 
\simeq \frac{ \mu_{\phi}}{b_j }
{\bf 1}_{n_j} \qquad \mbox{or} \qquad
M_{n_j} \simeq \frac{\alpha \Lambda^3}{\widetilde{N}_{c,j}} {\bf 1}_{n_j}
\nonu
\eea
corresponding to the nonzero $w$ coordinates of $n_j$ flavor D4-branes between 
the $D6_{-\theta,j}$-brane and the $NS5_j'$-brane.

One can think of  the brane configuration consisting of 
multiple outer NS5-branes as well as a single NS5'-brane,
$O6^{\pm}$-planes and eight D6-branes.
So we focus on the case where there is only a single outer 
NS5-brane \cite{LLL1,BHKL,EGKT,Ahn07-1} when we discuss the other brane
configurations containing this sub-brane configuration with NS5-brane,
NS5'-brane, $O6^{\pm}$-planes and eight D6-branes later. 

\section{$SU(N_c) \times SU(N_c')$ with $N_f$- and $N_f'$-fund. and bifund.}


\subsection{Electric theory}

The type IIA supersymmetric electric
brane configuration \cite{BH,Ahn07-3,AT97,Ahn07-8,Ahn07-9} corresponding to 
${\cal N}=1$ $SU(N_c) \times SU(N_c')$ gauge theory  with  
$N_f$-fundamental flavors $Q, \widetilde{Q}$,
$N_f'$-fundamental flavors $Q', \widetilde{Q}'$
and bifundamentals $X, \widetilde{X}$
can be described as follows: one middle NS5-brane, $2k'$
NS5'-branes, 
$N_c$- and $N_c'$-D4-branes, and $N_f$- and 
$N_f'$-D6-branes. 
The $N_c$-color D4-branes are suspended between 
the middle NS5-brane and the right NS5'-branes,
the $N_c'$-color D4-branes are suspended between 
the left NS5'-branes and the middle NS5-brane,
the $N_f$ D6-branes 
are located between the middle NS5-brane and the right NS5'-branes and
the $N_f'$ D6-branes 
are located between the left NS5'-branes and the middle NS5-brane.

Let us deform this theory
by adding the mass term 
and the quartic term for fundamental quarks.
The former can be achieved by displacing the D6-branes along $+v$
direction leading to their coordinates $v = + v_{D6}$ \cite{GK98} 
while the latter can be obtained by rotating the D6-branes
\cite{GK0710-1} 
by an angle 
$-\theta$ in $(w,v)$-plane and we denote them by $D6_{-\theta}$-branes. 
Then, in the electric gauge theory, the general 
deformed superpotential is
given by
\bea
W_{elec} & = & \frac{\alpha}{2} \tr (Q \widetilde{Q})^2 - m \tr Q
\widetilde{Q} 
 + \frac{\alpha'}{2} \tr (Q' \widetilde{Q}')^2 - m' \tr Q'
\widetilde{Q}' \nonu \\
& + & \left[ -\frac{\beta}{2}   \tr (X \widetilde{X})^{k'+1} + m_X
\tr X \widetilde{X}\right].
\label{elesuperpotential}
\eea 
The last two terms are due to the rotation of NS5'-branes where
$\beta=(\tan \omega_L + \tan \omega_R)$ 
and the relative displacement of D4-branes where the mass $m_X
=v_{NS5'}$
is the distance of D4-branes 
in $v$ direction.
When there are no D6-branes, this theory reduces 
to the one \cite{Ahn08-2} 
and the case where there exists further restriction on the 
number of NS5'-branes 
$k'=1$ has been studied in \cite{Ahn08-1two}.
We focus on the case with $k' \geq 2$ and the limit 
$\beta, m_X \rightarrow 0$.
When we take the Seiberg dual for the gauge group $SU(N_c)$, we put 
$\alpha'=0$ and $m'=0$(no displacement and no rotation of $N_f'$ D6-branes) 
while  for 
the Seiberg dual for the gauge group $SU(N_c')$, we take 
$\alpha=0$ and $m=0$.

Let us summarize the ${\cal N}=1$ supersymmetric electric brane
configuration with superpotential 
(\ref{elesuperpotential}) 
in type IIA string theory as follows and draw this in
Figure 7:

$\bullet$
One middle NS5-brane in $(012345)$ directions  with $w=0$

$\bullet$ 
$2k'$ NS5'-branes in  $(012389)$ directions $v=0$

$\bullet$
$N_f$ $D6_{-\theta}$-branes in (01237)
directions and
two other directions in $(v,w)$-plane

$\bullet$
$N_f'$ D6-branes in (0123789) directions 

$\bullet$
$N_c$- and $N_c'$-color D4-branes in $(01236)$ directions  with $v=0=w$ 

\begin{figure}[ht]
   \epsfxsize=2.0in 
\centerline{\epsffile{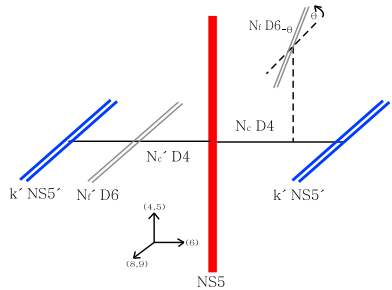}}
   \caption[FIG. \arabic{figure}.]{ 
The  ${\cal N}=1$ supersymmetric 
electric brane configuration for the gauge group $SU(N_c) \times SU(N_c')$ 
with bifundamentals $X, \widetilde{X}$ 
and fundamentals $Q, \widetilde{Q}, Q', \widetilde{Q}'$. 
Note that there exist multiple $2k'$ NS5'-branes.
A 
rotation of $N_f$ D6-branes in $(w,v)$-plane
corresponds to 
a quartic term for the fundamentals $Q, \widetilde{Q}$ while 
a displacement of $N_f$ D6-branes in $+v$ direction corresponds to a
mass term for the fundamentals $Q, \widetilde{Q}$.
}
\end{figure}

\subsection{Magnetic theory for $SU(N_c)$}

After we move a middle NS5-brane to the right all the way past the right 
NS5'-branes, we arrive at the Figure 8A.
Note that there exists a creation of $N_f$ D4-branes
connecting the $N_f$ $D6_{-\theta}$-branes 
and the $k'$ right NS5'-branes.
The linking number of NS5-brane from Figure 8A
is 
$
l_m = \frac{N_f}{2} -\widetilde{N}_c$.
On the other hand, the linking number of NS5-brane from Figure 7
is
$
l_e = -\frac{N_f}{2} + N_c -N_c'$. 
From these two relations, one obtains
the number of colors of dual magnetic theory as follows \cite{Ahn07-3}:
$
\widetilde{N}_c = N_f+N_c'-N_c$.

\begin{figure}[ht]
   \epsfxsize=3.5in 
\centerline{\epsffile{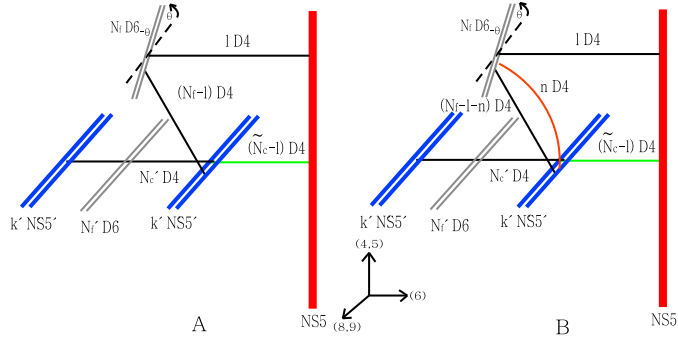}}
   \caption[FIG. \arabic{figure}.]{ 
The  ${\cal N}=1$ supersymmetric
magnetic brane configuration corresponding to Figure 7 with 
a misalignment between D4-branes when the gravitational potential of
the NS5-brane is ignored(8A) and nonsupersymmetric brane configuration
when  the gravitational potential of
the NS5-brane is considered(8B) . 
The $N_f$ flavor D4-branes connecting between
$D6_{-\theta}$-branes and NS5'-branes are splitting into $(N_f-l)$- and
$l$- D4-branes(8A). Further $n$- D4-branes among 
$(N_f-l)$- D4-branes are moved to the NS5-brane(8B). 
}
\end{figure}

The low energy theory on the color D4-branes 
has $SU(\widetilde{N}_c) \times SU(N_c')$ gauge group and  
$N_f$-fundamental dual quarks $q, \widetilde{q}$
coming from 4-4 strings connecting between the color $\widetilde{N}_c$ D4-branes and
$N_f$ flavor D4-branes as well as $Q', \widetilde{Q}', Y$ and
$\widetilde{Y}$ and gauge singlets.
Moreover, a magnetic meson field $M \equiv Q \widetilde{Q}$
is $N_f \times N_f$ matrix and comes from 
4-4 strings of $N_f$ flavor D4-branes.
Then the magnetic superpotential with the limit $\beta, m_X
\rightarrow 0$ 
is given by  
\bea
W_{dual} = \left[ \frac{1}{\Lambda} 
M q \widetilde{q} + Y \widetilde{F}' \widetilde{q} + 
\widetilde{Y} q F' + \Phi' Y \widetilde{Y} \right] 
+\frac{\alpha}{2} \tr M^2 - m M.
\label{dualW1}
\eea
The case with $k'=1$ and $\alpha=0$ was studied in \cite{Ahn07-3}.
Although the superpotential (\ref{dualW1}) 
does not depend on the multiplicity  $k'$ of
NS5'-branes in our particular limit, 
the difference from the previous result of \cite{Ahn07-3} appears as
two things: nonzero $\alpha$ in (\ref{dualW1}) and multiple NS5'-branes
in Figure 8.
Here other meson fields are given by 
$\Phi' \equiv X \widetilde{X},  
F' \equiv \widetilde{X} Q$ and 
$\widetilde{F}' \equiv X \widetilde{Q}$ \cite{Ahn07-3}.

For the supersymmetric vacua, one can compute the F-term equations for
this superpotential (\ref{dualW1}) 
and the expectation values for $M$ and $q \widetilde{q}$ 
are obtained.
The F-term equations are almost the same as the one in \cite{Ahn07-3}
and the derivative of (\ref{dualW1}) with respect 
to the meson field $M$ has $\alpha$ dependent term.
The vacuum expectation values for $Y, \widetilde{Y}, F'$ and
$\widetilde{F}'$
vanish as in \cite{Ahn07-3}.  

$\bullet$ Coincident $N_f$ $D6_{-\theta}$-branes and $k'$ NS5'-branes 

The theory has many nonsupersymmetric meta-stable ground states and 
when we rescale the meson field as
$
M = h \Lambda \Phi$ as before,
then the Kahler potential for $\Phi$ is canonical and the magnetic
quarks are canonical near the origin of field space \cite{ISS}.
Then the magnetic superpotential can be written as
\bea
W_{mag} = h \Phi  q   \widetilde{q} 
 +  
\frac{h^2 \mu_{\phi}}{2} \tr \Phi^2- h \mu^2 \tr \Phi + Y \widetilde{F}' \widetilde{q} + 
\widetilde{Y} q F' + \Phi' Y \widetilde{Y}
\nonu
\eea
where
$
\mu^2 = m \Lambda$ and  
$\mu_{\phi} = \alpha \Lambda^2$.
Now one splits 
the $(N_f-l) \times (N_f-l)$
block  at the lower right corner of $h\Phi$ and $q 
\widetilde{q}$ 
into blocks of 
size $n$ and $(N_f-l-n)$ as follows \cite{GK0710}:
\bea
h\Phi = \left(
\begin{array}{ccc}
0_l & 0 & 0  \\
0 & h \Phi_n & 0 \\
0 & 0 & \frac{\mu^2}{\mu_{\phi}} {\bf 1}_{N_f-l-n}
\end{array}
\right), \qquad
q  \widetilde{q} = \left(
\begin{array}{ccc}
\mu^2 {\bf 1}_l & 0 & 0  \\
0 & { \varphi}  \widetilde{\varphi}  &  0 \\
0 & 0 & 0_{N_f-l-n}
\end{array}
\right).
\nonu
\eea
Here $\varphi$ and $\widetilde{\varphi}$ are $n \times (\widetilde{N}_c-l)$
dimensional matrices and correspond to $n$ flavors of fundamentals of
the gauge group $SU(\widetilde{N}_c-l)$ which is unbroken.
In the brane configuration shown in Figure 8B, 
they correspond to 
fundamental strings connecting the $n$ flavor D4-branes and
$(\widetilde{N}_c-l)$
color D4-branes \cite{GK0710-1}.
The $\Phi_n$ and ${ \varphi} 
\widetilde{\varphi}$
are $n \times n$ matrices.
The supersymmetric ground state corresponds to
$h\Phi_n= \frac{\mu^2}{\mu_{\phi}} {\bf 1}_{n}$ and $ 
\varphi =0=\widetilde{\varphi}$ \footnote{Compared to a single gauge
group, there exist extra $k'$ left NS5'-branes, $N_f'$ D6-branes and $N_c'$ 
D4-branes. In Figure 8B, we consider the case where $k'$ left
NS5'-branes are far from $k'$ right NS5'-branes. From the result of
\cite{GKK}, in general, there exists a repulsive force between $N_c'$
D4-branes and $n$ D4-branes depending on their distance in $w$
direction and an rotation angle between those D4-branes. 
In order to weaken this effect, we need to take the limit where 
their distance in $w$
direction should be large and an rotation angle between them
should be small. Furthermore, the distance beteen $n$ D4-branes and
the NS5-brane should be small also.    }. 

Now the full one loop potential 
takes the form
\bea
\frac{V}{|h|^2}  =  
|\Phi_n  \varphi +  Y \widetilde{F}'|^2   
+  |\Phi_n \widetilde{\varphi} + F' \widetilde{Y}|^2
  +  
| \varphi  \widetilde{\varphi}-\mu^2 {\bf 1}_{n} + 
h \mu_{\phi} \Phi_n|^2 + b |h \mu|^2 \tr \Phi_n^{\dagger} \Phi_n, 
\nonu
\eea
where $b = \frac{(\ln 4-1)}{8\pi^2} \widetilde{N}_c$ and we do not
write down $\Phi_n$ or $\Phi_n^{\dagger}$-independent terms.
Differentiating this potential with respect to 
$\Phi_n^{\dagger}$ and putting $\varphi=0=
\widetilde{\varphi}$, one obtains
\bea
h \Phi_n 
\simeq \frac{ \mu_{\phi}}{b }
{\bf 1}_n \qquad \mbox{or} \qquad
M_n \simeq \frac{\alpha \Lambda^3}{\widetilde{N}_c} {\bf 1}_{n}
\label{sec5vac}
\eea
corresponding to the $w$ coordinates of $n$ curved flavor D4-branes between 
the $D6_{-\theta}$-branes and the NS5'-branes.

$\bullet$ Non-coincident $N_f$ $D6_{-\theta}$-branes and $k'$ NS5'-branes 

When the numbers of $D6_{-\theta}$-branes and 
the NS5'-branes are equal to each other $N_f=k'$,
we displace the $k'$ $D6_{-\theta}$-branes and NS5'-branes
given in Figure 8A
in the $v$ direction respectively
to two $k'$ different points denoted by 
$v_{D6_{-\theta,j}}$ and $v_{NS5_{j}'}$
where $j=1,2, \cdots, k'$, as in section 2.  
There are $
\sum_{j=1}^{k'} \widetilde{N}_{c,j}
= \widetilde{N}_c$ and $
\sum_{j=1}^{k'} N_{f,j}= N_f$.
Then the $n$ curved flavor D4-branes attached to them are displaced
also as $k'$ different $n_j$'s connecting between 
$D6_{-\theta,j}$-brane and
$NS5_{j}'$-brane.
Let us rescale the submeson field as
$M_{j} = h \Lambda \Phi_{j} $ \cite{Ahn08-2} and 
the Kahler potential for $\Phi_{j}$ is canonical and the magnetic
quarks $q_j$ and $\widetilde{q}_j$ 
are canonical near the origin of field space \cite{ISS}.
Then the magnetic superpotential 
can be rewritten in terms of $\Phi_j, q_j$ and $\widetilde{q}_j$ 
\bea
W_{mag} = \sum_{j=1}^{k'} 
\left[ h \Phi_j  q_j   \widetilde{q}_j 
 +  
\frac{h^2 \mu_{\phi}}{2} \tr \Phi_j^2- h \mu_j^2 \tr \Phi_j \right] + \cdots
\nonu
\eea
with
$
\mu_j^2 = m_j \Lambda_j$ and 
$\mu_{\phi} = \alpha \Lambda^2$ as before. 

One splits 
the $(N_{f,j}-l_j) \times (N_{f,j}-l_j)$
block  at the lower right corner of $h\Phi_j$ and $q_j 
\widetilde{q}_j$ 
into blocks of 
size $n_j$ and $(N_{f,j}-l_j-n_j)$ for all $j$ 
as follows \cite{Ahn08-2}:
\bea
h \Phi & = &  \left(
\begin{array}{ccccc}
0_{l+n} & 0 & 0 & \cdots  & 0 \\
0 & \frac{\mu_1^{2}}{\mu_{\phi}} 
{\bf 1}_{N_{f,1}- l_1-n_1} & 0 & \cdots & 0  \\
\cdot & \cdot & \cdot & \cdots & 0 \\
0 & 0 &  0 & 0 & \frac{\mu_{k'}^{2}}{\mu_\phi} 
{\bf 1}_{N_{f,k'}-l_{k'}-n_{k'}}  
\end{array}
\right) \nonu \\
&+ &  
\mbox{diag} (0_l, h \Phi_{n_1}, 
\cdots, h \Phi_{n_{k'}}, 0_{N_f-l-n})
\nonu
\eea
and
\bea
q   \widetilde{q}   & = &  \left(
\begin{array}{cccccc}
 \mu_1^2  {\bf 1}_{l_1} & 0 & 0 & \cdots & 0  & 0  \\
 \cdot  & \cdot & \cdot & \cdots & \cdot  & 0 \\
 0 & 0 & 0 & \cdots & \mu_{k'}^2  {\bf 1}_{l_{k'}}  & 0 \\ 
 0 & 0 & 0 & 0 & 0 & 0_{N_f-l} 
\end{array}
\right) \nonu \\
&+&
 \mbox{diag} (
0_l, \varphi_{n_1} 
\widetilde{\varphi}_{n_1}, 
\cdots,  \varphi_{n_{k'}}  
\widetilde{\varphi}_{n_{k'}}, 
0_{N_f-l-n})
\nonu
\eea
where $l =\sum_{j=1}^{k'} l_j$ and $n =\sum_{j=1}^{k'} n_j$.
Here 
$\varphi_{n_j}$ and $\widetilde{\varphi}_{n_j}$ are 
$n_j \times (\widetilde{N}_{c,j}-l_j)$
matrices and correspond to $n_j$-flavors of fundamentals of
the gauge group $SU(\widetilde{N}_{c,j}-l_j)$ which is unbroken.
The supersymmetric ground state corresponds to the vacuum expectation
values by
$h\Phi_{n_j}=\frac{\mu_j^2}{\mu_{\phi}}
{\bf 1}_{n_j}$ and $
\varphi_{n_j} \widetilde{\varphi}_{n_j}=0$.
The full one loop potential can be written similarly
and the local nonzero stable point arises as
\bea
h \Phi_{n_j} 
\simeq \frac{ \mu_{\phi}}{b_j }
{\bf 1}_{n_j} \qquad \mbox{or} \qquad
M_{n_j} \simeq \frac{\alpha \Lambda^3}{\widetilde{N}_{c,j}} {\bf 1}_{n_j}
\label{sec5vac1}
\eea
corresponding to the nonzero $w$ coordinates of $n_j$ flavor D4-branes between 
the $D6_{-\theta,j}$-brane and the $NS5_j'$-brane.
Then, the meta-stable states, for fixed $k'$ and $\theta$, 
are specified by the number of various D4-branes and the positions of
multiple $D6_{-\theta}$-branes and NS5'-branes. 

%

\section{
$Sp(N_c) \times SO(2N_c')$ with $N_f$-fund., $N_f'$-vectors.  
and bifund.}


\subsection{Electric theory}

The type IIA supersymmetric electric
brane configuration \cite{Ahn07-2,Tatar,Ahn97} corresponding to 
${\cal N}=1$ $Sp(N_c) \times SO(2N_c')$ gauge theory  with  
$N_f'$-vectors $Q'$,
$N_f$-fundamental flavors $Q$
and bifundamental $X$
can be described as follows: one middle NS5-brane, $2(2k'+1)$
NS5'-branes, 
$2N_c$- and $2N_c'$-D4-branes, and $2N_f$- and 
$2N_f'$-D6-branes as well as $O4^{\pm}$-planes. 
The $2N_c$-color D4-branes are suspended between 
the middle NS5-brane and the right NS5'-branes,
the $2N_c'$-color D4-branes are suspended between 
the left NS5'-branes and the middle NS5-brane,
the $2N_f$ D6-branes 
are located between the middle NS5-brane and the right NS5'-branes and
the $2N_f'$ D6-branes 
are located between the left NS5'-branes and the middle NS5-brane.

Let us deform this theory
by adding the mass term 
and the quartic term for fundamental quarks.
The former can be achieved by displacing the D6-branes along $ \pm v$
direction leading to their coordinates $v = \pm v_{D6}$ \cite{GK98} 
while the latter can be obtained by rotating the D6-branes
\cite{GK0710-1} 
by an angle 
$-\theta$ in $(w,v)$-plane and we denote them by $D6_{-\theta}$-branes. 
Then, in the electric gauge theory, the general deformed superpotential is
given by
\bea
W_{elec} & = & \frac{\alpha}{2} \tr (Q Q)^2 - m \tr Q Q +
 \frac{\alpha'}{2} \tr (Q' Q')^2 - m' \tr Q'
Q' \nonu \\
  &+& \left[  -
\frac{\beta}{2} \tr (X X)^{2k'+2} + m_X \tr X X\right].
\label{esuperpotent}
\eea 
The last two terms are due to the rotation of NS5'-branes where
$\beta=\tan \omega $ 
and the relative displacement of D4-branes where the mass $m_X =v_{NS5'}$ 
is the distance of D4-branes in $v$ direction.
When there are no D6-branes, this theory reduces to the one \cite{Ahn08-2} 
and the case where there exists a further restriction on 
the number of NS5'-branes $k'=0$ 
has been studied in \cite{Ahn08-1two}.
We focus on the case with $k' \geq 1$ and 
$\beta, m_X \rightarrow 0$.
When we take the Seiberg dual for the gauge group $Sp(N_c)$, we put 
$\alpha'=0$ and $m'=0$
while  for 
the Seiberg dual for the gauge group $SO(2N_c')$, we take 
$\alpha=0$ and $m=0$ instead.

Let us summarize the ${\cal N}=1$ supersymmetric electric brane
configuration with superpotential 
(\ref{esuperpotent}) 
in type IIA string theory as follows and draw this in
Figure 9:

$\bullet$
One middle NS5-brane in $(012345)$ directions  with $w=0$

$\bullet$ 
$2(2k'+1)$ NS5'-branes in  $(012389)$ directions $v=0$

$\bullet$
$2N_f$ $D6_{-\theta}$-branes in (01237)
directions and
two other directions in $(v,w)$-plane

$\bullet$
$2N_f'$ D6-branes in (0123789) directions 

$\bullet$
$2N_c$- and $2N_c'$-color D4-branes in $(01236)$ directions  with $v=0=w$ 

$\bullet$ $O4^{\pm}$-planes in (01236) directions with $w=0=v$

\begin{figure}[ht]
   \epsfxsize=2.0in 
\centerline{\epsffile{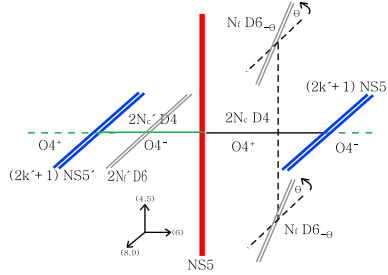}}
   \caption[FIG. \arabic{figure}.]{ 
The  ${\cal N}=1$ supersymmetric 
electric brane configuration for the gauge group $Sp(N_c) \times SO(2N_c')$ 
with bifundamental $X$ 
and fundamentals and vectors $Q, Q'$. 
Note that there are multiple $2(2k'+1)$ NS5'-branes.
A 
rotation of $N_f$ D6-branes in $(w,v)$-plane
corresponds to 
a quartic term for the fundamentals $Q$ while 
a displacement of $N_f$ D6-branes in $ \pm v$ direction corresponds to a
mass term for the fundamentals $Q$.
}
\end{figure}

\subsection{Magnetic theory for $Sp(N_c)$}

After we move  the middle NS5-brane to the right all the way past
the right NS5'-branes, the linking number
of NS5-brane from Figure 10A is given by 
$
l_m =\frac{(2N_f)}{2}-1-(1)-2\widetilde{N}_c
$.
Originally, it was given by
$
l_e=-\frac{(2N_f)}{2}+ 1 -(-1)
+2N_c-2N_c'
$
from Figure 9.
Therefore, by the linking number conservation and equating these two
$l$'s each other, 
we are left with the number of colors in the magnetic
theory \cite{Ahn07-2}
$
\widetilde{N}_c = N_f+N_c'-N_c-2$.

\begin{figure}[ht]
   \epsfxsize=3.0in 
\centerline{\epsffile{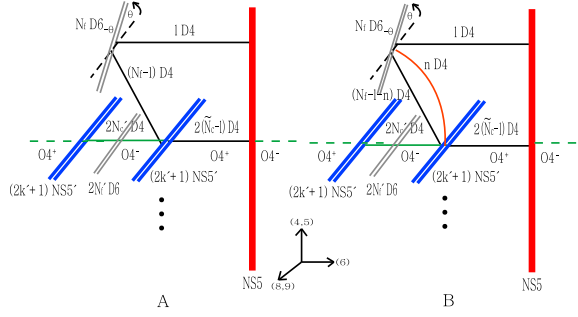}}
   \caption[FIG. \arabic{figure}.]{ 
The  ${\cal N}=1$ supersymmetric
magnetic brane configuration corresponding to Figure 9 with 
a misalignment between D4-branes when the gravitational potential of
the NS5-brane is ignored(10A) and nonsupersymmetric brane configuration
when  the gravitational potential of
the NS5-brane is considered(10B) . 
The $N_f$ flavor D4-branes connecting between
$D6_{-\theta}$-branes and NS5'-branes are splitting into $(N_f-l)$- and
$l$- D4-branes(10A). Further $n$- D4-branes among 
$(N_f-l)$- D4-branes are moved to the NS5-brane(10B). 
}
\end{figure}

The low energy theory on the color D4-branes 
has $Sp(\widetilde{N}_c) \times SO(2N_c')$ gauge group and  
$N_f$-fundamental dual quarks $q$
coming from 4-4 strings connecting between the color $\widetilde{N}_c$ D4-branes and
$N_f$ flavor D4-branes as well as $Q', Y$ and gauge singlets.
Moreover, a magnetic meson field $M \equiv Q Q$
is $2N_f \times 2N_f$ matrix and comes from 
4-4 strings of $2N_f$ flavor D4-branes.
Then the magnetic superpotential with the limit $\beta, m_X
\rightarrow 0$ 
is given by  
\bea
W_{dual} = \left[ \frac{1}{\Lambda} M q q  + Y \Phi'
Y + q N Y \right]
+ \frac{\alpha}{2} \tr M^2 - m M.
\label{finalsuper}
\eea
The case with $k'=0$ and $\alpha=0$ was studied in \cite{Ahn07-2}.
Although the superpotential (\ref{finalsuper}) 
does not depend on the multiplicity  $(2k'+1)$ of
NS5'-branes in our particular limit, 
the difference from the previous result of \cite{Ahn07-2} appears as
1) nonzero $\alpha$ in (\ref{finalsuper}) and 2) multiple NS5'-branes
in Figure 10A.
Here other meson fields are given by 
$\Phi' \equiv  X X$ and $ N \equiv  Q X$.

One can compute the F-term equations for
this superpotential (\ref{finalsuper}) 
and the expectation values for $M$ and $q q$ 
are obtained.
The F-term equations are almost the same as the one in \cite{Ahn07-2}
and the derivative of (\ref{finalsuper}) with respect 
to the meson field $M$ has $\alpha$-dependent term.
The vacuum expectation values for $Y$ and
$N$
vanish as in \cite{Ahn07-2}.  

$\bullet$ Coincident $N_f$ $D6_{-\theta}$-branes and $(k'+\frac{1}{2})$ NS5'-branes 

The theory has many nonsupersymmetric meta-stable ground states and 
when we rescale the meson field as
$
M = h \Lambda \Phi$ as before.
Then the magnetic superpotential can be written as
\bea
W_{mag} = h \Phi  q  q 
 +  
\frac{h^2 \mu_{\phi}}{2} \tr \Phi^2- h \mu^2 \tr \Phi 
 + Y \Phi'
Y + q N Y
\nonu
\eea
where
$
\mu^2 = m \Lambda$ and  
$\mu_{\phi} = \alpha \Lambda^2$.
Now one splits 
the $2(N_f-l) \times 2(N_f-l)$
block  at the lower right corner of $h\Phi$ and $q q$ 
into blocks of 
size $2n$ and $2(N_f-l-n)$ as follows \cite{GK0710}:
\bea
h\Phi = \left(
\begin{array}{ccc}
0_{2l} & 0 & 0  \\
0 & h \Phi_{2n}  & 0 \\
0 & 0 & \frac{\mu^2}{\mu_{\phi}} {\bf 1}_{N_f-l-n} \otimes i \sigma_2
\end{array}
\right), \qquad
q  q = \left(
\begin{array}{ccc}
\mu^2 {\bf 1}_{2l} & 0 & 0  \\
0 &  \varphi  \varphi  &  0 \\
0 & 0 & 0_{2(N_f-l-n)}
\end{array}
\right).
\nonu
\eea
Here $\varphi$ is $2n \times 2(\widetilde{N}_c-l)$
dimensional matrices and correspond to $2n$ flavors of fundamentals of
the gauge group $Sp(\widetilde{N}_c-l)$ which is unbroken.
The $\Phi_{2n}$ and ${ \varphi} \varphi$
are $2n \times 2n$ matrices.
The supersymmetric ground state corresponds to
$h\Phi_{2n}= \frac{\mu^2}{\mu_{\phi}} {\bf 1}_{n} \otimes i \sigma_2$ and $ 
\varphi =0$. 

Now the full one loop potential 
takes the form
\bea
\frac{V}{|h|^2}  =  
|\Phi_{2n}  \varphi +  N Y |^2   
  +  
| \varphi \varphi-\mu^2 {\bf 1}_{2n} + 
h \mu_{\phi} \Phi_{2n}|^2 + b |h \mu|^2 \tr \Phi_{2n} \Phi_{2n}, 
\nonu
\eea
where $b = \frac{(\ln 4-1)}{8\pi^2} \widetilde{N}_c$ and we do not
write down $\Phi_{2n}$-independent terms.
Differentiating this potential with respect to 
$\Phi_{2n}$ and putting $\varphi=0$, one obtains
\bea
h \Phi_{2n} 
\simeq \frac{ \mu_{\phi}}{b }
{\bf 1}_n \otimes i \sigma_2 \qquad \mbox{or} \qquad
M_{2n} \simeq \frac{\alpha \Lambda^3}{\widetilde{N}_c} {\bf 1}_{n}
\otimes i \sigma_2
\nonu
\eea
corresponding to the $w$ coordinates of $n$ curved flavor D4-branes between 
the $D6_{-\theta}$-branes and the NS5'-branes.

$\bullet$ Non-coincident $N_f$ $D6_{-\theta}$-branes and 
$(k'+\frac{1}{2})$ NS5'-branes 

When the numbers of $D6_{-\theta}$-branes and
the number of NS5'-branes minus one are equal $N_f=k'$,
we displace the $k'$ upper $D6_{-\theta}$-branes and upper NS5'-branes, 
given in Figure 10A, in the $+v$ direction respectively
to two $k'$ different points denoted by 
$v_{D6_{-\theta,j}}$ and $v_{NS5_{j}'}$
where $j=1,2, \cdots, k'$(and their mirrors in the $-v$ direction).
A single  NS5'-brane is located at $v=0$.
Then there exist 
$
\sum_{j=1}^{k'} \widetilde{N}_{c,j}(\equiv N_{f,j}-N_{c,j}-2)
= \widetilde{N}_c$ and $
\sum_{j=1}^{k'} N_{f,j} = N_f$.
When all the upper $NS5_{j}'$-branes and 
$D6_{-\theta,j}$-branes($j=1, 2, \cdots, k'$) 
are distinct, the low energy  
physics corresponds to $k'$ decoupled supersymmetric 
gauge theories with gauge groups 
$
\left[\prod_{j=1}^{k'} Sp( \widetilde{N}_{c,j})\right] \times SO(2N_c')$.

One deforms the Figure 10B by displacing the multiple
$D6_{-\theta}$-branes and $NS5'$-branes 
along $v$ direction \cite{Ahn08-2}.
Then the $n$ curved flavor D4-branes attached to them are displaced
also as $k'$ different $n_j$'s connecting between 
$D6_{-\theta,j}$-brane and
$NS5_{j}'$-brane.
Then the magnetic superpotential
can be rewritten in terms of $\Phi_j, q_j$
\bea
W_{mag} = \sum_{j=1}^{k'} 
\left[ h \Phi_j  q_j  q_j 
 +  
\frac{h^2 \mu_{\phi}}{2} \tr \Phi_j^2- h \mu_j^2 \tr \Phi_j \right] + \cdots
\nonu
\eea
with
$
\mu_j^2 = m_j \Lambda_j$ and 
$\mu_{\phi} = \alpha \Lambda^2$ as before. 

One splits 
the $2(N_{f,j}-l_j) \times 2(N_{f,j}-l_j)$
block  at the lower right corner of $h\Phi_j$ and $q_j
q_j$ 
into blocks of 
size $2n_j$ and $2(N_{f,j}-l_j-n_j)$ for all $j$ 
as follows:
\bea
h \Phi & = &  \left(
\begin{array}{ccccc}
0_{2(l+n)} & 0 & 0 & \cdots  & 0 \\
0 & \frac{\mu_1^{2}}{\mu_{\phi}} 
{\bf 1}_{N_{f,1}- l_1-n_1} \otimes i \sigma_2 & 0 & \cdots & 0  \\
\cdot & \cdot & \cdot & \cdots & 0 \\
0 & 0 &  0 & 0 & \frac{\mu_{k'}^{2}}{\mu_\phi} 
{\bf 1}_{N_{f,k'}-l_{k'}-n_{k'}} \otimes i \sigma_2 
\end{array}
\right) \nonu \\
&+&  
\mbox{diag} (0_{2l}, h \Phi_{2n_1}, 
\cdots, h \Phi_{2n_{k'}}, 0_{2(N_f-l-n)})
\nonu
\eea
and
\bea
q  q   & = &  \left(
\begin{array}{cccccc}
 \mu_1^2  {\bf 1}_{2l_1} & 0 & 0 & \cdots & 0  & 0  \\
 \cdot  & \cdot & \cdot & \cdots & \cdot  & 0 \\
 0 & 0 & 0 & \cdots & \mu_{k'}^2  {\bf 1}_{2l_{k'}}  & 0 \\ 
 0 & 0 & 0 & 0 & 0 & 0_{2(N_f-l)} 
\end{array}
\right) \nonu \\
&+&
 \mbox{diag} (
0_{2l}, \varphi_{2n_1} 
\varphi_{2n_0}, 
\cdots,  \varphi_{2n_{k'}} \varphi_{2n_{k'}}, 
0_{2(N_f-l-n)})
\nonu
\eea
where $l =\sum_{j=1}^{k'} l_j$ and $n =\sum_{j=1}^{k'} n_j$.
Here 
$\varphi_{2n_j}$  is 
$2n_j \times 2(\widetilde{N}_{c,j}-l_j)$
matrices and correspond to $2n_j$-flavors of fundamentals of
the gauge group $Sp(\widetilde{N}_{c,j}-l_j)$ which is unbroken.
Moreover,
the $\Phi_{2n_j}$ and ${ \varphi}_{2n_j} \varphi_{2n_j}$
are $2n_j \times 2n_j$ matrices.
The supersymmetric ground state corresponds to the vacuum expectation
values by
$h\Phi_{2n_j}=\frac{\mu_j^2}{\mu_{\phi}}
{\bf 1}_{n_j} \otimes i \sigma_2$ and $
\varphi_{2n_j}=0$.
The full one loop potential can be written similarly
and the local nonzero stable point arises as
\bea
h \Phi_{2n_j} 
\simeq \frac{\mu_{\phi}}{b_j}
{\bf 1}_{n_j} \otimes i \sigma_2 \qquad \mbox{or} \qquad
M_{2n_j} \simeq \frac{\alpha  
\Lambda^3}{\widetilde{N}_{c,j}}  {\bf 1}_{n_j} \otimes i \sigma_2
\nonu
\eea
corresponding to the $w$ coordinates of $n_j$ flavor D4-branes between 
the $D6_{-\theta,j}$-brane and the $NS5_j'$-brane(and their mirrors).
Then, the meta-stable states, for fixed $k'$ and $\theta$, 
are specified by the number of various D4-branes and the positions of
multiple $D6_{-\theta}$-branes and NS5'-branes. 

%

\section{
$SU(N_c) \times SO(2N_c')$ with $N_f$-fund., $2N_f'$-vectors  
and bifund.}

\subsection{Electric theory}

The type IIA supersymmetric electric
brane configuration \cite{LO,Ahn07-3} corresponding to 
${\cal N}=1$ $SU(N_c) \times SO(2N_c')$ gauge theory  with  
$N_f$-fundamental flavors $Q, \widetilde{Q}$,
$2N_f'$-vectors $Q'$
and bifundamentals $X, \widetilde{X}$
can be described as follows: two NS5-branes, $2k'$
NS5'-branes, 
$N_c$- and $2N_c'$-D4-branes, $2N_f$- and $2N_f'$-D6-branes 
and $O6^{+}$-plane. 
The $2N_c'$-color D4-branes are suspended between 
the two NS5-branes and
the $N_c$-color D4-branes are suspended between 
the NS5-brane and the NS5'-branes(and their mirrors).

Let us deform this theory
by adding the mass term 
and the quartic term for the fundamentals.
The former can be achieved by displacing the D6-branes along $+v$
direction leading to their coordinates $v = + v_{D6}$ \cite{GK98} 
while the latter can be obtained by rotating the D6-branes
\cite{GK0710-1} 
by an angle 
$-\theta$ in $(w,v)$-plane and we denote them by $D6_{-\theta}$-branes. 
Then, in the electric gauge theory, the general deformed superpotential is
given by
\bea
W_{elec} & = & 
 \frac{\alpha}{2} \tr (Q \widetilde{Q})^2 - m \tr Q \widetilde{Q}  +
 \frac{\alpha'}{2} \tr (Q' Q')^2 - m' \tr Q'
Q' \nonu \\
&+& \left[-\frac{\beta}{2}   \tr (X \widetilde{X})^{k'+1} +
 m_X \tr X
\widetilde{X} \right]. 
\label{superelec7}
\eea
The last two terms are due to the rotation of NS5'-branes where
$\beta=\tan \omega$ 
and the relative displacement of D4-branes where the mass $m_X =v_{NS5'}$ 
is the distance in $v$ direction.
This reduces to the one \cite{Ahn07-3} when 
$k'=1$ and $\alpha=0=\alpha'=m'$.
We focus on the case with $k' \geq 2$ and 
$\beta, m_X \rightarrow 0$.
When we take the Seiberg dual for the gauge group $SU(N_c)$, we put 
$\alpha'=0$ and $m'=0$.

Let us summarize the ${\cal N}=1$ supersymmetric electric brane
configuration with superpotential 
(\ref{superelec7}) 
in type IIA string theory as follows and draw this in
Figure 11:

$\bullet$
Two NS5-branes in $(012345)$ directions  with $w=0$

$\bullet$ 
$2k'$ NS5'-branes in  $(012389)$ directions $v=0$

$\bullet$
$N_f$ $D6_{\pm \theta}$-branes in (01237)
directions and
two other directions in $(v,w)$-plane

$\bullet$
$2N_f'$ D6-branes in (0123789) directions 

$\bullet$
$N_c$- and $2N_c'$-color D4-branes in $(01236)$ directions  with $v=0=w$ 

$\bullet$ $O6^{+}$-plane in (0123789) directions with $x^6=0=v$ 

\begin{figure}[ht]
   \epsfxsize=3.5in 
\centerline{\epsffile{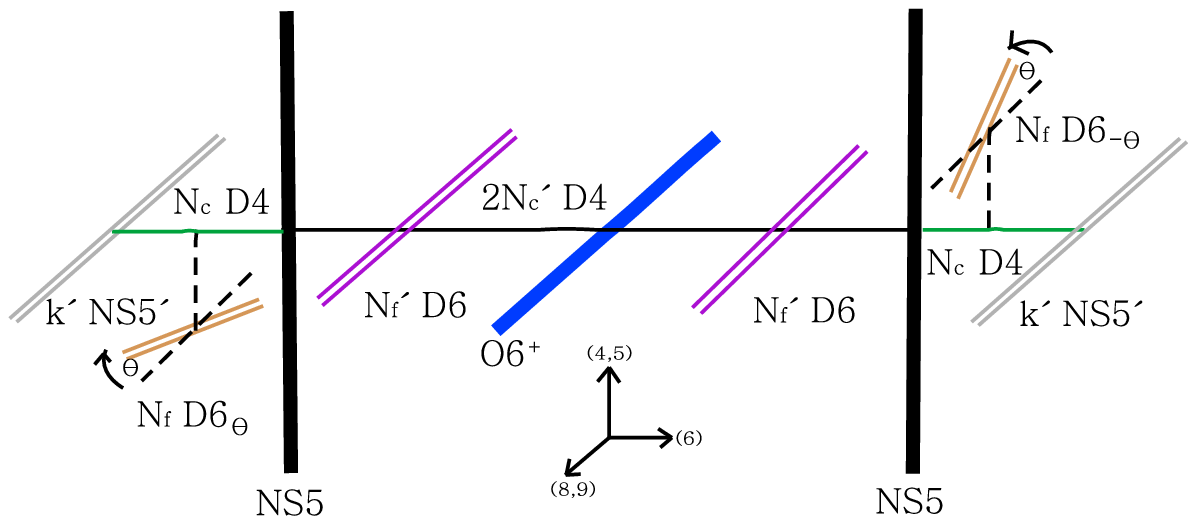}}
   \caption[FIG. \arabic{figure}.]{ 
The  ${\cal N}=1$ supersymmetric 
electric brane configuration for the gauge group $SU(N_c) \times SO(2N_c')$ 
with bifundamental $X, \widetilde{X}$, 
fundamentals $Q, \widetilde{Q}$ and vectors $Q'$. 
Note that there are multiple $k'$ NS5'-branes(and its mirrors).
A 
rotation of $N_f$ D6-branes in $(w,v)$-plane
corresponds to 
a quartic term for the fundamentals while 
a displacement of $N_f$ D6-branes in $ \pm v$ direction corresponds to a
mass term for the fundamentals.
}
\end{figure}

\subsection{Magnetic theory for $SU(N_c)$}

After we move the right NS5-brane to the right all the way past the right 
NS5'-branes, we arrive at the Figure 12A.
Note that there exists a creation of $N_f$ D4-branes
connecting the $N_f$ $D6_{-\theta}$-branes and the $k'$ right NS5'-branes.
The linking number of NS5-brane from Figure 12A
is 
$
l_m = \frac{N_f}{2} -\widetilde{N}_c$.
On the other hand, the linking number of NS5-brane from Figure 11
is
$
l_e = -\frac{N_f}{2} + N_c -N_c'$. 
From these two relations, one obtains
the number of colors of dual magnetic theory 
as follows \cite{Ahn07-3}:
$
\widetilde{N}_c = N_f+N_c'-N_c$.

\begin{figure}[ht]
   \epsfxsize=3.5in 
\centerline{\epsffile{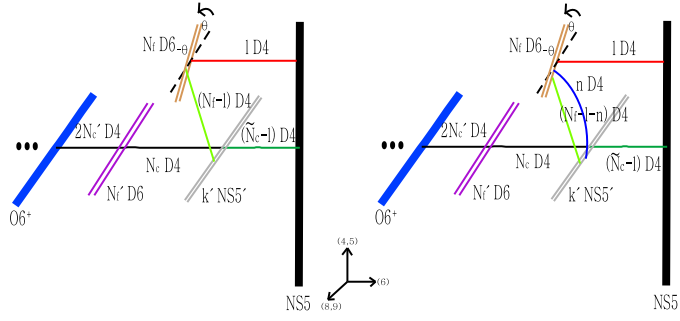}}
   \caption[FIG. \arabic{figure}.]{ 
The  ${\cal N}=1$ supersymmetric
magnetic brane configuration corresponding to Figure 11 with 
a misalignment between D4-branes when the gravitational potential of
the NS5-brane is ignored(12A) and nonsupersymmetric brane configuration
when  the gravitational potential of
the NS5-brane is considered(12B) . 
The $N_f$ flavor D4-branes connecting between
$D6_{-\theta}$-branes and NS5'-branes are splitting into $(N_f-l)$- and
$l$- D4-branes(12A). Further $n$- D4-branes among 
$(N_f-l)$- D4-branes are moved to the NS5-brane(12B). 
}
\end{figure}

The low energy theory on the color D4-branes 
has $SU(\widetilde{N}_c) \times SO(2N_c')$ gauge group and  
$N_f$-fundamental dual quarks $q, \widetilde{q}$
coming from 4-4 strings connecting between the color $\widetilde{N}_c$ D4-branes and
$N_f$ flavor D4-branes as well as $Q', Y, \widetilde{Y}$ and gauge singlets.
Moreover, a magnetic meson field $M \equiv Q \widetilde{Q}$
is $N_f \times N_f$ matrix and comes from 
4-4 strings of $N_f$ flavor D4-branes.
Then the magnetic superpotential with the limit $\beta, m_X
\rightarrow 0$ 
is given by  
\bea
W_{dual} = \left[ \frac{1}{\Lambda} 
M q \widetilde{q} + Q' \Phi' Q'  + Y \widetilde{F}' \widetilde{q} + 
\widetilde{Y} q F' + \Phi' Y \widetilde{Y} \right] 
+\frac{\alpha}{2} \tr M^2 - m M.
\label{dualW11}
\eea
The case with $k'=1$ and $\alpha=0$ was studied in \cite{Ahn07-3}.
Although the superpotential (\ref{dualW11}) 
does not depend on the multiplicity  $k'$ of
NS5'-branes in our particular limit, 
the difference from the previous result of \cite{Ahn07-3} appears as
two things: nonzero $\alpha$ in (\ref{dualW11}) and multiple NS5'-branes
in Figure 12.
Here other meson fields are given by 
$\Phi' \equiv X \widetilde{X},  
F' \equiv \widetilde{X} Q$ and 
$\widetilde{F}' \equiv X \widetilde{Q}$ \cite{Ahn07-3}.

For the supersymmetric vacua, one can compute the F-term equations for
this superpotential (\ref{dualW11}) 
and the expectation values for $M$ and $q \widetilde{q}$ 
are obtained.
The F-term equations are almost the same as the one in \cite{Ahn07-3}
and the derivative of (\ref{dualW11}) with respect 
to the meson field $M$ has $\alpha$ dependent term.
The vacuum expectation values for $Y, \widetilde{Y}, F',
\widetilde{F}'$ and $Q'$
vanish as in \cite{Ahn07-3}.  

$\bullet$ Coincident $N_f$ $D6_{-\theta}$-branes and $k'$ NS5'-branes 

By following the description of subsection 5.2, 
one obtains the local nonzero stable point given by (\ref{sec5vac})
corresponding to the $w$ coordinates of $n$ curved flavor D4-branes between 
the $D6_{-\theta}$-branes and the NS5'-branes in Figure 12B.

$\bullet$ Non-coincident $N_f$ $D6_{-\theta}$-branes and $k'$ NS5'-branes 

The local nonzero stable point arises as (\ref{sec5vac1})
corresponding to the nonzero $w$ coordinates of $n_j$ flavor D4-branes between 
the $D6_{-\theta,j}$-brane and the $NS5_j'$-brane.


%
%
%

\section{
$SU(N_c) \times SU(N_c')$ with a symm. and
bifund. }


\subsection{Electric theory}

The type IIA supersymmetric electric
brane configuration \cite{Ahn07-5,Ahn07-6,Ahn07-7,Ahn07-10} corresponding to 
${\cal N}=1$ $SU(N_c) \times SU(N_c')$ gauge theory  with  
a symmetric tensor $S$, a conjugate symmetric tensor $\widetilde{S}$
and bifundamentals $X, \widetilde{X}$
can be described as follows: $(2k+1)$ NS5-brane, $2k'$
NS5'-branes, 
$N_c$- and $N_c'$-D4-branes and $O6^{+}$-plane. 
The $N_c$-color D4-branes are suspended between 
the left NS5'-branes  and the right NS5'-branes and
the $N_c'$-color D4-branes are suspended between 
the right NS5'-branes and the right NS5-branes(and their mirrors).

Let us deform this theory
by adding the mass term 
and the higher order term for the bifundamentals.
The former can be achieved by displacing the NS5-branes along $+v$
direction leading to their coordinates $v = + v_{NS5}$ \cite{GK98} 
while the latter can be obtained by rotating the NS5-branes
\cite{GK0710-1} 
by an angle 
$-\theta$ in $(w,v)$-plane and we denote them by $NS5_{-\theta}$-branes. 
Then, in the electric gauge theory, the general 
deformed superpotential by including the mass term and higher order
term for symmetric tensor matter is
given by
\bea
W_{elec} = -\frac{\alpha}{2}   \tr (X \widetilde{X})^{k+1} +
 m \tr X
\widetilde{X} +\left[
-\frac{\beta}{2}   \tr (S \widetilde{S})^{k'+1}  
+ m_S \tr S \widetilde{S} \right]. 
\label{superelec}
\eea
The last two terms are due to the rotation of NS5'-branes where
$\beta=\tan \omega$ 
and the relative displacement of D4-branes where the mass $m_S =v_{NS5'}$ 
is the distance in $v$ direction.
This reduces to the one \cite{Ahn08-1two} when 
$k=k'=1$ and $\beta=m_S=0$.
We focus on the case with $k, k' \geq 2$ and 
$\beta, m_S \rightarrow 0$.

Let us summarize the ${\cal N}=1$ supersymmetric electric brane
configuration with superpotential 
(\ref{superelec}) 
in type IIA string theory as follows and draw this in
Figure 13:

$\bullet$ One NS5-brane in (012345) directions with $w=0=x^6$

$\bullet$ $2k'$ NS5'-branes in (012389) directions with $v=0$

$\bullet$ $k$ $NS5_{\pm \theta}$-branes in (012389) directions with $v=0$

$\bullet$ $N_c$- and $N_c'$-color D4-branes in (01236) directions with $v=0=w$

$\bullet$ $O6^{+}$-plane in (0123789) directions with $x^6=0=v$ 

\begin{figure}[ht]
   \epsfxsize=2.0in 
\centerline{\epsffile{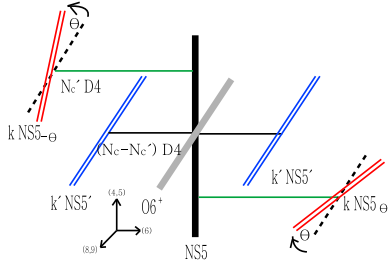}}
   \caption[FIG. \arabic{figure}.]{ 
The  ${\cal N}=1$ supersymmetric 
electric brane configuration for the gauge group $SU(N_c) \times SU(N_c')$ 
with a symmetric tensor $S, \widetilde{S}$ 
and bifundamentals $X, \widetilde{X}$. 
Note that there are multiple $k$ $NS5_{\pm \theta }$-branes and $2k'$ NS5'-branes. 
A 
rotation of $k$ NS5-branes in $(w,v)$-plane
corresponds to 
a higher term for the bifundamentals while 
a displacement of $k$ NS5-branes in $ \pm v$ direction corresponds to a
mass term for the bifundamentals.
}
\end{figure}

\subsection{Magnetic theory for $SU(N_c)$}

We apply the Seiberg dual to the 
$SU(N_c)$ factor and the two $k'$ NS5'-branes are 
interchanged each other. 
The linking number of the right NS5'-brane in Figure 14A is 
$l_m =
-\frac{\widetilde{N}_c}{k'}+  
\frac{N_c'}{k'}$
and the linking number of the left NS5'-brane in Figure 13 is given by 
$l_e=
\frac{N_c}{k'}-\frac{N_c'}{k'}$.
This leads to the fact that the number of D4-branes becomes
$
\widetilde{N}_c = 2 N_c'-N_c$. 

\begin{figure}[ht]
   \epsfxsize=4.0in 
\centerline{\epsffile{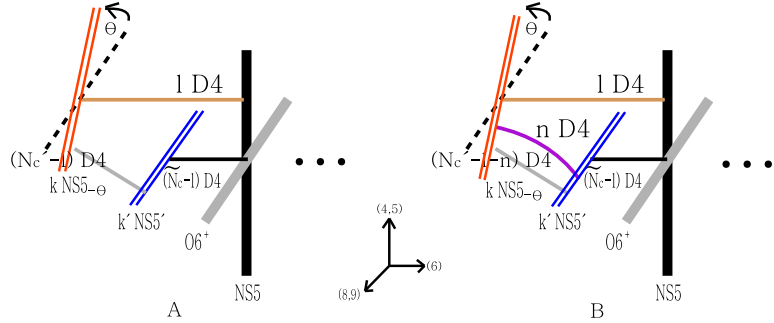}}
   \caption[FIG. \arabic{figure}.]{ 
The  ${\cal N}=1$ supersymmetric
magnetic brane configuration corresponding to Figure 13 with 
a misalignment between D4-branes when the gravitational potential of
the NS5-brane is ignored(13A) and nonsupersymmetric brane configuration
when  the gravitational potential of
the NS5-brane is considered(13B) . 
The $N_c'$ flavor D4-branes connecting between
$NS5_{-\theta}$-branes and NS5'-branes are splitting into $(N_c'-l)$- and
$l$- D4-branes(14A). Further $n$- D4-branes among 
$(N_c'-l)$- D4-branes are moved to the NS5-brane(14B).
}
\end{figure}

The low energy theory on the color D4-branes 
has $SU(\widetilde{N}_c) \times SU(N_c')$ gauge group and  
$N_c'$-fundamental dual ``quarks'' $Y, \widetilde{Y}$
coming from 4-4 strings connecting between the color $\widetilde{N}_c$ D4-branes and
$N_c'$ flavor D4-branes as well as $s$ and $\widetilde{s}$.
Moreover, a single magnetic meson field $M \equiv X \widetilde{X}$
is $N_c' \times N_c'$ matrix and comes from 
4-4 strings of $N_c'$ flavor D4-branes.
Then the magnetic superpotential with the limit $\beta, m_S
\rightarrow 0$ 
is given by  
\bea
W_{dual} =  \frac{1}{\Lambda} M Y \widetilde{Y}- \frac{\alpha}{2} M^{k+1}  + m M. 
\label{sup}
\eea
The case with $k=1=k'$  and $\alpha=0$ was studied in \cite{Ahn07-5}.
Although the superpotential (\ref{sup}) 
does not depend on the multiplicity  $k'$ of
NS5'-branes in our particular limit, 
the difference from the previous result of \cite{Ahn07-5} appears as
two things: nonzero $\alpha$ in (\ref{sup}) which has explicit $k$
dependent term and multiple $NS5_{-\theta}$-branes and NS5'-branes
in Figure 14A.
For the supersymmetric vacua, one can compute the F-term equations for
this superpotential (\ref{sup}) 
and the expectation values for $M$ and $Y \widetilde{Y}$ 
are obtained.
The F-term equations are the same as the one in \cite{Ahn08-2}.

$\bullet$ Coincident $k$ $NS5_{-\theta}$-branes and $k'$ NS5'-branes 

The theory has many nonsupersymmetric meta-stable ground states
due to the fact that there exists an attractive
gravitational interaction
between the flavor D4-branes and the NS5-brane from the DBI action.
When we rescale the meson field as
$M = h \Lambda \Phi $,
then the Kahler potential for $\Phi$ is canonical and the magnetic
quarks $Y$ and $\widetilde{Y}$ 
are canonical near the origin of field space.
Then the magnetic superpotential (\ref{sup}) 
can be rewritten as
\bea
W_{dual} = 
 h \Phi  Y   \widetilde{Y} 
 +  
\frac{ \mu_{\phi}}{2} h^{k+1} \tr \Phi^{k+1}- h \mu^2 \tr \Phi 
\label{Dual}
\eea
with the new couplings
$
\mu^2 = -m \Lambda$ and  
$\mu_{\phi} = -\alpha \Lambda^{k+1}$.

Now one splits 
the $(N_c'-l) \times (N_c'-l)$
block  at the lower right corner of $M$ and $Y
\widetilde{Y}$ 
into blocks of 
size $n$ and $(N_c'-l-n)$ and then they are 
rewritten as follows \cite{GK0710}:
\bea
h\Phi = \left(
\begin{array}{ccc}
0{\bf 1}_l & 0 & 0  \\
0 & h \Phi_n & 0 \\
0 & 0 & \left[\frac{2\mu^2}{\mu_{\phi}(k+1)}\right]^{\frac{1}{k}} 
{\bf 1}_{N_c'-l-n}
\end{array}
\right), \qquad
Y  \widetilde{Y} = \left(
\begin{array}{ccc}
\mu^2 {\bf 1}_l & 0 & 0  \\
0 & { \varphi}  \widetilde{\varphi}  &  0 \\
0 & 0 & 0_{N_c'-l-n}
\end{array}
\right).
\label{Eigen}
\eea
Here $\varphi$ and $\widetilde{\varphi}$ are $n \times (\widetilde{N}_c-l)$
matrices and correspond to $n$-flavors of fundamentals of
the gauge group $SU(\widetilde{N}_c-l)$ which is unbroken.
The supersymmetric ground state corresponds to the vacuum expectation
values by
$h\Phi_n=\left[\frac{2\mu^2}{\mu_{\phi}(k+1)}\right]^{\frac{1}{k}}
{\bf 1}_{n}$ and $
\varphi \widetilde{\varphi} =0$. 

The full one loop potential for 
$\Phi_n, \varphi$, and $\widetilde{\varphi}$
from (\ref{Dual}) and (\ref{Eigen}) including the one loop result
\cite{ISS} takes the form
\bea
\frac{V}{|h|^2}  =  
|\Phi_n  \varphi|^2   
+  |\Phi_n  \widetilde{\varphi}  |^2
  +  
\left| \varphi \widetilde{\varphi}-\mu^2 {\bf 1}_{n} + 
\frac{(k+1)h^k}{2} 
\mu_{\phi}  \Phi_n^k \right|^2 + b |h \mu|^2 \tr \Phi_n^{\dagger} \Phi_n, 
\label{pot}
\eea
where the positive numerical constant $b$ is given by
$b = \frac{(\ln 4-1)}{8\pi^2} \widetilde{N}_c$.
Differentiating this potential (\ref{pot}) with respect to 
$\Phi_n^{\dagger}$ and putting $\varphi=0=
\widetilde{\varphi}$, one obtains
\bea
h \Phi_n 
\simeq \left[ \frac{2b }{k(k+1) \mu_{\phi}} \right]^{\frac{1}{k-2}}
{\bf 1}_n \qquad \mbox{or} \qquad
M_n \simeq \left[
\frac{\widetilde{N}_c}{\alpha k(k+1) 
\Lambda^3} \right]^{\frac{1}{k-2}} {\bf 1}_{n}.
\label{vac}
\eea
It is evident that 
the $(N_c'-l-n)$ flavor D4-branes between 
the $NS5_{-\theta}$-branes and 
the NS5'-branes are related to the corresponding eigenvalues 
of $h\Phi$ (\ref{Eigen}), i.e.,  
$ 
\left[\frac{2\mu^2}{\mu_{\phi}(k+1)}\right]^{\frac{1}{k}} 
{\bf 1}_{N_c'-l-n}$ and 
the intersection point between the 
$(N_c'-l-n)$ D4-branes and the NS5'-branes is also given 
by $(v, w)=(0, +v_{NS5_{-\theta}} \cot \theta)$.

$\bullet$ Non-coincident $k$ $NS5_{-\theta}$-branes and $k'$ NS5'-branes 

When the numbers of $NS5_{-\theta}$-branes and 
the NS5'-branes are equal to each other $k=k'$,
we displace the $k'$ $NS5_{-\theta}$-branes and NS5'-branes
given in Figure 14A
in the $v$ direction respectively
to two $k'$ different points denoted by 
$v_{NS5_{-\theta,j}}$ and $v_{NS5_{j}'}$
where $j=1,2, \cdots, k'$, as in section 2.  
There are $
\sum_{j=1}^{k'} \widetilde{N}_{c,j}
= \widetilde{N}_c$ and $
\sum_{j=1}^{k'} N_{c,j}' = N_c'$.

One deforms the Figure 14B by displacing the multiple
$NS5_{-\theta}$-branes and $NS5'$-branes 
along $v$ direction.
Then the $n$ curved flavor D4-branes attached to them are displaced
also as $k'$ different $n_j$'s connecting between 
$NS5_{-\theta,j}$-brane and
$NS5_{j}'$-brane.
Let us rescale the submeson field as
$M_{j} = h \Lambda \Phi_{j} $ \cite{Ahn08-2} and 
the Kahler potential for $\Phi_{j}$ is canonical and the magnetic
quarks $Y_j$ and $\widetilde{Y}_j$ 
are canonical near the origin of field space \cite{ISS}.
Then the magnetic superpotential 
can be rewritten in terms of $\Phi_j, Y_j$ and $\widetilde{Y}_j$ 
\bea
W_{mag} = \sum_{j=1}^{k} 
\left[ h \Phi_j  Y_j   \widetilde{Y}_j 
 +  
\frac{h^{k+1} \mu_{\phi}}{2} \tr \Phi_j^{k+1}- h \mu_j^2 \tr \Phi_j \right]
\nonu
\eea
with
$
\mu_j^2 = -m_j \Lambda_j$ and 
$\mu_{\phi} = -\alpha \Lambda^{k+1}$ as before. 

One splits 
the $(N_{c,j}'-l_j) \times (N_{c,j}'-l_j)$
block  at the lower right corner of $h\Phi_j$ and $Y_j 
\widetilde{Y}_j$ 
into blocks of 
size $n_j$ and $(N_{c,j}'-l_j-n_j)$ for all $j$ 
as follows \cite{Ahn08-2}:
\bea
h \Phi & = & \left[\frac{2}{\mu_{\phi}(k+1)}\right]^{\frac{1}{k}} \left(
\begin{array}{ccccc}
0_{l+n} & 0 & 0 & \cdots  & 0 \\
0 & \mu_1^{\frac{2}{k}} 
{\bf 1}_{N_{c,1}'-l_1-n_1} & 0 & \cdots & 0  \\
\cdot & \cdot & \cdot & \cdots & 0 \\
0 & 0 &  0 & 0 & \mu_k^{\frac{2}{k}} 
{\bf 1}_{N_{c,k}'-l_{k}-n_k}  
\end{array}
\right) \nonu \\
& + & 
\mbox{diag} (0_{l}, h \Phi_{n_1}, \cdots, h \Phi_{n_k},
0_{N_c'-l-n})
\nonu
\eea
and
\bea
Y  \widetilde{Y}   & = &  \left(
\begin{array}{cccccc}
\mu_1^2  {\bf 1}_{l_1} & 0 & 0 & \cdots & 0  & 0  \\
\cdot  & \cdot & \cdot & \cdots & \cdot  & 0 \\
0 & 0 & 0 & \cdots & \mu_k^2  {\bf 1}_{l_k}  & 0 \\ 
0 & 0 & 0 & 0 & 0 & 0_{N_c'-l} 
\end{array}
\right) \nonu \\
& + &   \mbox{diag} (
0_{l}, \varphi_{n_1} \widetilde{\varphi}_{n_1}, 
\cdots,  \varphi_{n_k} \widetilde{\varphi}_{n_k}, 
0_{N_c'-l-n})
\nonu
\eea
where $l =\sum_{j=1}^{k'} l_j$ and $n =\sum_{j=1}^{k'} n_j$.
Here 
$\varphi_{n_j}$ and $\widetilde{\varphi}_{n_j}$ are 
$n_j \times (\widetilde{N}_{c,j}-l_j)$
matrices and correspond to $n_j$-flavors of fundamentals of
the gauge group $SU(\widetilde{N}_{c,j}-l_j)$ which is unbroken.
The supersymmetric ground state corresponds to the vacuum expectation
values by
$h\Phi_{n_j}=
\left[\frac{2\mu_j^2}{\mu_{\phi}(k+1)}\right]^{\frac{1}{k}}
{\bf 1}_{n_j}$ and $
\varphi_{n_j} \widetilde{\varphi}_{n_j}=0$.
The full one loop potential can be written similarly
and the local nonzero stable point arises as
\bea
h \Phi_{n_j} 
\simeq \left[ \frac{2b_j }{k(k+1) \mu_{\phi}} \right]^{\frac{1}{k-2}}
{\bf 1}_{n_j} \qquad \mbox{or} \qquad
M_{n_j} \simeq \left[
\frac{\widetilde{N}_{c,j}}{\alpha k(k+1) 
\Lambda_j^3} \right]^{\frac{1}{k-2}} {\bf 1}_{n_j}
\nonu
\eea
corresponding to the nonzero $w$ coordinates of $n_j$ flavor D4-branes between 
the $NS5_{-\theta,j}$-brane and the $NS5_j'$-brane.
Then, the meta-stable states, for fixed $k, k'$ and $\theta$, 
are specified by the number of various D4-branes and the positions of
multiple $NS5_{-\theta}$-branes and NS5'-branes. 

%

\section{
$SU(N_c) \times SU(N_c')$ with an antisymm., 
eight-fund. and bifund.}

\subsection{Electric theory}

The type IIA supersymmetric electric
brane configuration \cite{Ahn07-5} corresponding to 
${\cal N}=1$ $SU(N_c) \times SU(N_c')$ gauge theory  with  
an antisymmetric tensor $A$, a conjugate symmetric tensor
$\widetilde{S}$,
eight fundamentals $\hat{Q}$
and bifundamentals $X, \widetilde{X}$
can be described as follows: Two NS5-branes, $(2k'+1)$
NS5'-branes, 
$N_c$- and $N_c'$-D4-branes, eight D6-branes and $O6^{\pm}$-planes. 
The $N_c$-color D4-branes are suspended between 
the left NS5-brane and the right NS5-brane  and
the $N_c'$-color D4-branes are suspended between 
the right NS5-brane and the right NS5'-branes(and their mirrors).

Let us deform this theory
by adding the mass term 
and the higher order term for the bifundamentals.
The former can be achieved by displacing the NS5'-branes along $+v$
direction leading to their coordinates $v = + v_{NS5'}$ \cite{GK98} 
while the latter can be obtained by rotating the NS5'-branes
\cite{GK0710-1} 
by an angle 
$-\theta$ in $(w,v)$-plane and we denote them by $NS5_{-\theta}$-branes. 
Then, in the electric gauge theory, the general deformed 
superpotential is
given by
\bea
W_{elec} =  
 -\frac{\alpha}{2}   \tr (X \widetilde{X})^{k'+1} +
 m \tr X
\widetilde{X} +\left[
-\frac{\beta}{2}   \tr (A \widetilde{S})^{2}  
 +
\hat{Q} \widetilde{S} \hat{Q} \right].
\label{electrics1}
\eea 
The third term is due to the rotation of NS5-branes where
$\beta=\tan \omega$.
This theory reduces to the one \cite{Ahn08-1two} when 
$k'=1$ and $\beta=0$.
We focus on the case with $k' \geq 2$ and 
$\beta \rightarrow 0$.

Let us summarize the ${\cal N}=1$ supersymmetric electric brane
configuration with superpotential 
(\ref{electrics1}) 
in type IIA string theory as follows and draw this in
Figure 15:

$\bullet$ Two NS5-branes in (012345) directions with $w=0$

$\bullet$ One NS5'-brane in (012389) directions with $v=0=x^6$

$\bullet$ $k'$ $NS5_{\pm \theta}$-branes in (012345) directions with $w=0$

$\bullet$ $N_c$- and $N_c'$-color D4-branes in (01236) directions with $v=0=w$

$\bullet$ Eight half D6-branes in (0123789) directions with $x^6=0=v$

$\bullet$ $O6^{\pm}$-planes in (0123789) directions with $x^6=0=v$

\begin{figure}[ht]
   \epsfxsize=2.0in 
\centerline{\epsffile{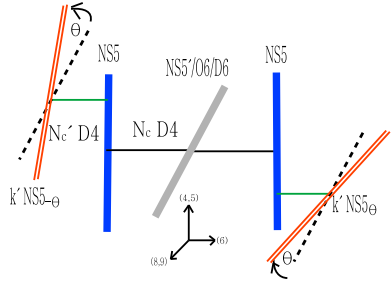}}
   \caption[FIG. \arabic{figure}.]{ 
The  ${\cal N}=1$ supersymmetric 
electric brane configuration for the gauge group $SU(N_c) \times SU(N_c')$ 
with an antisymmetric tensor $A, \widetilde{S}$, eight fundamentals $\hat{Q}$, 
and bifundamentals $X, \widetilde{X}$. 
Note that there are $k'$ $NS5_{\pm \theta}$-branes. A 
rotation of $k'$ NS5'-branes in $(w,v)$-plane
corresponds to 
a higher term for the bifundamentals while 
a displacement of $k'$ NS5-branes in $ \pm v$ direction corresponds to a
mass term for the bifundamentals.
}
\end{figure}

\subsection{Magnetic theory for $SU(N_c)$}

We apply the Seiberg dual to the 
$SU(N_c)$ factor and the two NS5-branes are 
interchanged each other. 
The linking number of the right NS5-brane in Figure 16A is 
$l_m =
\frac{4}{2}
-\widetilde{N}_c+  N_c'$
and the linking number of the left NS5-brane in Figure 15 is given by 
$l_e=
-\frac{4}{2} +
N_c-N_c'$.
This leads to the fact that the number of D4-branes becomes \cite{Ahn07-5}
$
\widetilde{N}_c = 2 N_c'-N_c+4$.

\begin{figure}[ht]
   \epsfxsize=3.5in 
\centerline{\epsffile{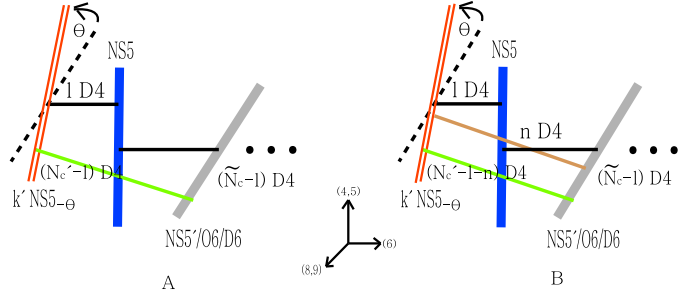}}
   \caption[FIG. \arabic{figure}.]{ 
The  ${\cal N}=1$ supersymmetric
magnetic brane configuration corresponding to Figure 15 with 
a misalignment between D4-branes when the gravitational potential of
the NS5-brane is ignored(16A) and nonsupersymmetric brane configuration
when  the gravitational potential of
the NS5-brane is considered(16B) . 
The $N_c'$ flavor D4-branes connecting between
$NS5_{-\theta}$-branes and NS5'-brane are splitting into $(N_c'-l)$- and
$l$- D4-branes(16A). Further $n$- D4-branes among 
$(N_c'-l)$- D4-branes are moved to the NS5-brane(16B). 
}
\end{figure}

The low energy theory on the color D4-branes 
has $SU(\widetilde{N}_c) \times SU(N_c')$ gauge group and  
$N_c'$-fundamental dual ``quarks'' $Y, \widetilde{Y}$
coming from 4-4 strings connecting between the color $\widetilde{N}_c$ D4-branes and
$N_c'$ flavor D4-branes as well as $a, \widetilde{s}$ and $\hat{q}$.
Moreover, a single magnetic meson field $M \equiv X \widetilde{X}$
is $N_c' \times N_c'$ matrix and comes from 
4-4 strings of $N_c'$ flavor D4-branes.
Then the magnetic superpotential with the limit $\beta \rightarrow 0$ 
is given by  
\bea
W_{dual} =  \frac{1}{\Lambda} M Y \widetilde{Y}- \frac{\alpha}{2} M^{k'+1}
 + m M + \hat{q}
 \widetilde{s} \hat{q}. 
\label{super}
\eea
The case with $k'=1$  and $\alpha=0$ was studied in \cite{Ahn07-5}.
The difference from the previous result of \cite{Ahn07-5} appears as
two things: nonzero $\alpha$ in (\ref{super}) which has explicit $k'$
dependent term and multiple $NS5_{\mp\theta}$-branes
in Figure 16A.
For the supersymmetric vacua, one can compute the F-term equations for
this superpotential (\ref{super}) 
and the expectation values for $M$ and $Y \widetilde{Y}$ 
are obtained.

$\bullet$ Coincident $k'$ $NS5_{-\theta}$-branes  

One obtains the vacuum expectation value (\ref{vac}) with a
replacement $k$ by $k'$ by applying the prescription of  subsection 8.2. 
This provides the $w$ coordinates of $n$ flavor D4-branes between 
the $NS5_{-\theta}$-branes and the NS5'-brane in Figure 16B.

$\bullet$ Non-coincident $k'$ $NS5_{-\theta}$-branes  

These  non-coincident $NS5_{-\theta}$-branes can be obtained by taking those
quark masses being unequal. Then all the previous descriptions for the
meta-stable states can be applied in this
case also without any difficulty. 

\section{$SU(N_c) \times SU(N_c')$ with $N_f$- and $N_f'$-fund., a symm. 
and bifund.}


\subsection{Electric theory}

The type IIA supersymmetric electric
brane configuration \cite{Ahn07-4} corresponding to 
${\cal N}=1$ $SU(N_c) \times SU(N_c')$ gauge theory  with  
$N_f$-fundamental flavors $Q, \widetilde{Q}$,
$N_f'$-fundamental flavors $Q', \widetilde{Q}'$, a symmetric tensor 
$S$, a conjugate symmetric tensor $\widetilde{S}$
and bifundamentals $X, \widetilde{X}$
can be described as follows: Three NS5-branes, $2k'$
NS5'-branes, 
$N_c$- and $N_c'$-D4-branes, and $2N_f$- and 
$2N_f'$-D6-branes and $O6^{+}$-plane. 
The $N_c$-color D4-branes are suspended between 
the left NS5'-branes and the right NS5'-branes,
the $N_c'$-color D4-branes are suspended between 
the right NS5'-branes and the right NS5-brane(and their mirrors),
the $N_f$ D6-branes 
are located between the middle NS5-brane and the right NS5'-branes and
the $N_f'$ D6-branes 
are located between the right NS5'-branes and the right NS5-brane(and
their mirrors).

Let us deform this theory
by adding the mass term 
and the quartic term for fundamental quarks.
The former can be achieved by displacing the D6-branes along $ \pm v$
direction leading to their coordinates $v = \pm v_{D6}$ \cite{GK98} 
while the latter can be obtained by rotating the D6-branes
\cite{GK0710-1} 
by an angle 
$ \mp \theta$ in $(w,v)$-plane and we denote them 
by $D6_{\mp \theta}$-branes(or $D6_{\mp \theta'}$-branes for the
second gauge group). 
Then, in the electric gauge theory, the deformed superpotential is
given by
\bea
W_{elec} & = & \frac{\alpha}{2} \tr (Q \widetilde{Q})^2 - m \tr Q
\widetilde{Q} +
 \frac{\alpha'}{2} \tr (Q' \widetilde{Q}')^2 - m' \tr Q'
\widetilde{Q}' \nonu \\
& + & \left[-\frac{\beta}{2}   \tr (S \widetilde{S})^{k'+1} + m_S
\tr S \widetilde{S}\right]
 +\left[-\frac{\gamma}{2}   \tr (X \widetilde{X})^{2} +
 m_X \tr X
\widetilde{X} \right].
\label{potentialsome}
\eea 
The last four terms are due to the rotation of NS5'-branes and NS5-branes where
$\beta=\tan \omega'$ and $\gamma =\tan \omega$ 
and the relative displacement of D4-branes where the mass $m_S=v_{NS5'}$ 
and the mass $m_X =v_{NS5}$ 
are the distance in $v$ direction.
The case of $k'=1$ and $\alpha=0$ with $\beta=\gamma=m_S=m_X=0$
was studied in \cite{Ahn07-4}.
We focus on the case with $k' \geq 2$ and 
$\beta, \gamma, m_S, m_X \rightarrow 0$.
When we take the Seiberg dual for the gauge group $SU(N_c)$, we put 
$\alpha'=0$ and $m'=0$
and  for 
the Seiberg dual on the gauge group $SU(N_c')$ we take 
$\alpha=0$ and $m=0$.

Let us summarize the ${\cal N}=1$ supersymmetric electric brane
configuration with superpotential 
(\ref{potentialsome}) 
in type IIA string theory as follows and draw this in
Figure 17:

$\bullet$ 
$2k'$ NS5'-branes in  $(012389)$ directions $v=0$

$\bullet$ 
Three NS5-branes in  $(012345)$ directions $w=0$

$\bullet$
$N_f$ $D6_{\pm \theta}$-branes in (01237)
directions and
two other directions in $(v,w)$-plane

$\bullet$
$N_f'$ $D6_{\pm \theta'}$-branes in (01237) directions 
 and
two other directions in $(v,w)$-plane

$\bullet$
$N_c$- and $N_c'$-color D4-branes in $(01236)$ directions  with $v=0=w$ 

$\bullet$ $O6^{+}$-plane in (0123789) directions with $x^6=0=v$ 

\begin{figure}[ht]
   \epsfxsize=3.5in 
\centerline{\epsffile{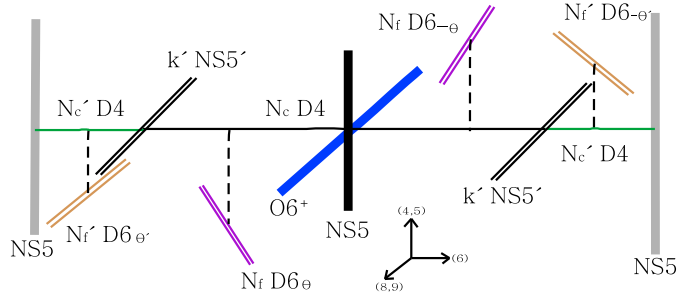}}
   \caption[FIG. \arabic{figure}.]{ 
The  ${\cal N}=1$ supersymmetric 
electric brane configuration for the gauge group $SU(N_c) \times SU(N_c')$ 
with a symmetric tensor $S, \widetilde{S}$, 
bifundamentals $X, \widetilde{X}$ and fundamentals $Q, \widetilde{Q},
Q', \widetilde{Q}'$. 
Note that there are $2k'$ NS5'-branes.
A 
rotation of $N_f(N_f')$ D6-branes in $(w,v)$-plane
corresponds to 
a quartic term for the fundamentals $Q, \widetilde{Q}(Q', \widetilde{Q}')$ while 
a displacement of $N_f(N_f')$ D6-branes in $ \pm v$ direction corresponds to a
mass term for the fundamentals $Q, \widetilde{Q}(Q', \widetilde{Q}')$.
}
\end{figure}

\subsection{Magnetic theory for $SU(N_c)$}

After the left NS5'-branes and $D6_{\theta}$-branes and 
the right NS5'-branes and $D6_{-\theta}$-branes 
are exchanged each other, we arrive at the Figure 18A.
One reads off 
the number of colors of dual magnetic theory from section 4 by
noticing the new $N_c'$ dependence:
$
\widetilde{N}_c = 2(N_f+N_c')-N_c+4(k'-1)$.

\begin{figure}[ht]
   \epsfxsize=4.0in 
\centerline{\epsffile{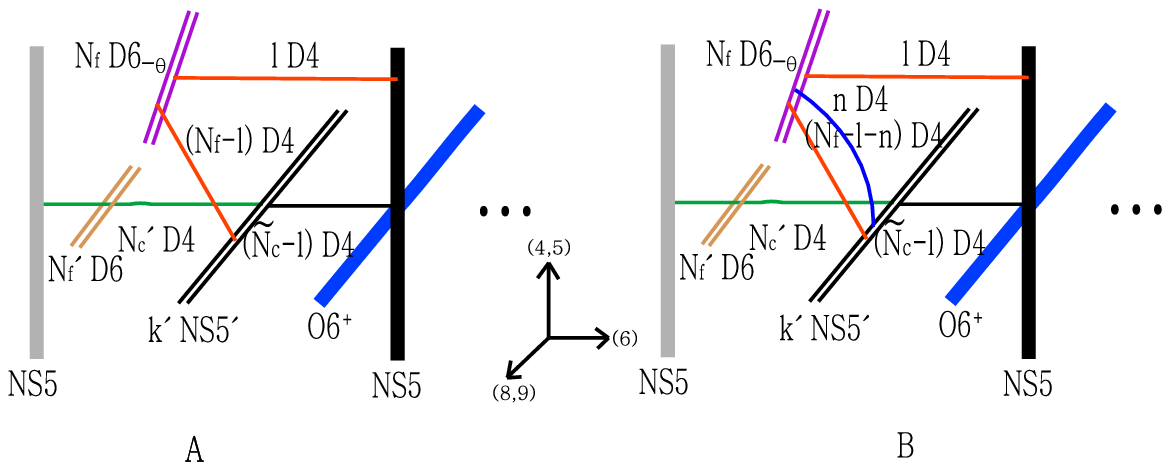}}
   \caption[FIG. \arabic{figure}.]{ 
The  ${\cal N}=1$ supersymmetric
magnetic brane configuration corresponding to Figure 17 with 
a misalignment between D4-branes when the gravitational potential of
the NS5-brane is ignored(18A) and nonsupersymmetric brane configuration
when  the gravitational potential of
the NS5-brane is considered(18B) . 
The $N_f$ flavor D4-branes connecting between
$D6_{-\theta}$-branes and NS5'-branes are splitting into $(N_f-l)$- and
$l$- D4-branes(18A). Further $n$- D4-branes among 
$(N_f-l)$- D4-branes are moved to the NS5-brane(18B). 
}
\end{figure}

The low energy theory on the color D4-branes 
has $SU(\widetilde{N}_c) \times SU(N_c')$ gauge group and  
$N_f$-fundamental dual quarks $q, \widetilde{q}$
coming from 4-4 strings connecting between the color $\widetilde{N}_c$ D4-branes and
$N_f$ flavor D4-branes as well as $Q', \widetilde{Q}', s,
\widetilde{s}, Y$ and $\widetilde{Y}$ and gauge singlets.
Moreover, a magnetic meson field $M \equiv Q \widetilde{Q}$
is $N_f \times N_f$ matrix and comes from 
4-4 strings of $N_f$ flavor D4-branes.
Then the magnetic superpotential with the limit $\beta, \gamma, m_S, m_X
\rightarrow 0$ 
is given by  
\bea
W_{dual} = \left[ \frac{1}{\Lambda} 
M q \widetilde{s} s \widetilde{q} + Y \widetilde{F}' \widetilde{q} + 
\widetilde{Y} q F' + \Phi' Y \widetilde{Y} \right] 
+\frac{\alpha}{2} \tr M^2 - m M.
\label{dualW2}
\eea
The case with $k'=1$ and $\alpha=0$ was studied in \cite{Ahn07-4} as
mentioned before.
Although the superpotential (\ref{dualW2}) 
does not depend on the multiplicity  $k'$ of
NS5'-branes in our particular limit, 
the difference from the previous result of \cite{Ahn07-4} appears as
1) nonzero $\alpha$ in (\ref{dualW2}) and 2) mutiple NS5'-branes
in Figure 18A.
Here other meson fields are given by 
$\Phi' \equiv X \widetilde{X},  
F' \equiv \widetilde{X} Q$ and 
$\widetilde{F}' \equiv X \widetilde{Q}$ \cite{Ahn07-4}.

For the supersymmetric vacua, one can compute the F-term equations for
this superpotential (\ref{dualW2}) 
and the expectation values for $M$ and $q \widetilde{s} s \widetilde{q}$ 
are obtained.
The F-term equations are almost the same as the one in \cite{Ahn07-4}
and the derivative of (\ref{dualW2}) with respect 
to the meson field $M$ has $\alpha$ dependent term.
The vacuum expectation values for $Y, \widetilde{Y}, F'$ and
$\widetilde{F}'$
vanish as in \cite{Ahn07-4}.  

$\bullet$ Coincident $N_f$ $D6_{-\theta}$-branes and $k'$ NS5'-branes 

One gets the local stable point (\ref{sec4vac})
corresponding to the $w$ coordinates of $n$ curved flavor D4-branes between 
the $D6_{-\theta}$-branes and the NS5'-branes in Figure 18B by following the
prescription of subsection 4.2.

$\bullet$ Non-coincident $N_f$ $D6_{-\theta}$-branes and $k'$ NS5'-branes 

The local nonzero stable point arises similarly.

\subsection{Magnetic theory for $SU(N_c')$}

After we move the right NS5'-branes to the right all the way 
past the right 
NS5-brane(and its mirrors to the left), we arrive at the Figure 19A.
Note that there exists a creation of $k' N_f'$ D4-branes
connecting the $N_f'$ $D6_{-\theta'}$-branes and the NS5'-brane.
The linking number of NS5'-brane from Figure 19A
is 
$
l_m = \frac{N_f'}{2} -\frac{\widetilde{N}_c}{k'}$.
On the other hand, the linking number of NS5'-brane from Figure 17
is
$
l_e = -\frac{N_f'}{2} + \frac{N_c'}{k'} -\frac{N_c}{k'}$. 
From these two relations, one obtains
the number of colors of dual magnetic theory
$
\widetilde{N}_c' = k' N_f' + N_c - N_c'$.
In this subsection we consider only the single number of NS5'-brane
$
k' =1
$
in order to deal with a single meson field.

\begin{figure}[ht]
   \epsfxsize=4.5in 
\centerline{\epsffile{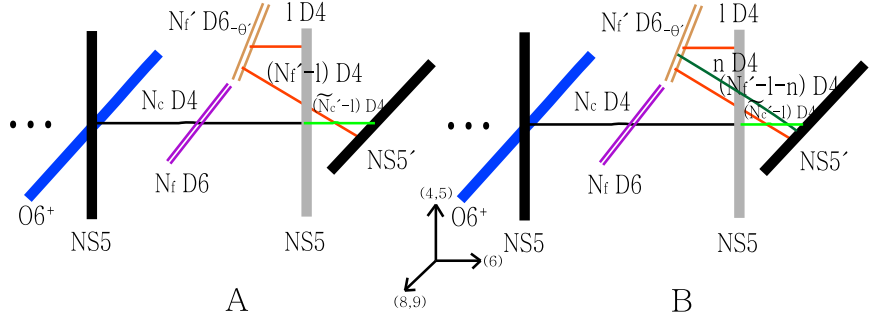}}
   \caption[FIG. \arabic{figure}.]{ 
The  ${\cal N}=1$ supersymmetric
magnetic brane configuration corresponding to Figure 18 with 
a misalignment between D4-branes when the gravitational potential of
the NS5-brane is ignored(19A) and nonsupersymmetric brane configuration
when  the gravitational potential of
the NS5-brane is considered(19B). 
The $N_f'$ flavor D4-branes connecting between
$D6_{-\theta'}$-branes and NS5'-brane are splitting into $(N_f'-l)$- and
$l$- D4-branes(19A). Further $n$- D4-branes among 
$(N_f'-l)$- D4-branes are moved to the NS5-brane(19B). 
}
\end{figure}

The low energy theory on the color D4-branes 
has $SU(N_c) \times SU(\widetilde{N}_c')$ gauge group and  
$N_f'$-fundamental dual quarks $q', \widetilde{q}'$
coming from 4-4 strings connecting between the color $\widetilde{N}_c'$ D4-branes and
$N_f'$ flavor D4-branes as well as $Q, \widetilde{Q}, S,
\widetilde{S}, Y, \widetilde{Y}$ and gauge singlets.
Moreover, a magnetic meson field $M \equiv Q' \widetilde{Q}'$
is $N_f' \times N_f'$ matrix and comes from 
4-4 strings of $N_f'$ flavor D4-branes.
Then the magnetic superpotential with the limit $\beta, \gamma, m_S, m_X
\rightarrow 0$ 
is given by  
\bea
W_{dual} = \left[ \frac{1}{\Lambda} M q' \widetilde{q}'    + 
Y F q' +
\widetilde{Y} \widetilde{q}' \widetilde{F} + \Phi' Y \widetilde{Y} \right] 
+\frac{\alpha}{2} \tr M^2 - m' M.
\label{dual10}
\eea
The case with $k'=1$ and $\alpha=0$ was studied in \cite{Ahn07-4}.
Although the superpotential (\ref{dual10}) 
does not depend on the multiplicity  $k'$ of
NS5'-branes in our particular limit, 
the difference from the previous result of \cite{Ahn07-4} appears as
nonzero $\alpha$ in (\ref{dual10}).
Here other meson fields are given by 
$\Phi' \equiv X \widetilde{X},  
F \equiv X Q'$ and 
$\widetilde{F} \equiv \widetilde{X} \widetilde{Q}'$ \cite{Ahn07-4}.

For the supersymmetric vacua, one can compute the F-term equations for
this superpotential (\ref{dual10}) 
and the expectation values for $M$ and $q' \widetilde{q}'$ 
are obtained.
The F-term equations are almost the same as the one in \cite{Ahn07-4}
and the derivative of (\ref{dual10}) with respect 
to the meson field $M$ has $\alpha$ dependent term.
The vacuum expectation values for $Y, \widetilde{Y}, F$ and
$\widetilde{F}$
vanish as in \cite{Ahn07-4}.  

$\bullet$ Coincident $N_f'$ $D6_{-\theta'}$-branes  

One obtains the local stable point as (\ref{sec5vac}) by changing the
role of $Q$ and $\widetilde{Q}$ into the one of $Q'$ and $\widetilde{Q}'$.
This gives the $w$ coordinates of $n$ flavor D4-branes between 
the $D6_{-\theta'}$-branes and the NS5'-brane in Figure 19B.

$\bullet$ Non-coincident $N_f'$ $D6_{-\theta'}$-branes 

These  non-coincident $D6_{-\theta}$-branes can be obtained by taking those
quark masses being unequal. Then all the previous descriptions for the
meta-stable states can be applied in this
case also without any difficulty. 

%

\section{$SU(N_c) \times SU(N_c')$ with $N_f$- and $N_f'$-fund.,
an antisymm., eight-fund.  
and bifund.}


\subsection{Electric theory}

The type IIA supersymmetric electric
brane configuration \cite{Ahn07-4} corresponding to 
${\cal N}=1$ $SU(N_c) \times SU(N_c')$ gauge theory  with  
$N_f$-fundamental flavors $Q, \widetilde{Q}$,
$N_f'$-fundamental flavors $Q', \widetilde{Q}'$, an antisymmetric tensor 
$A$, a conjugate symmetric tensor $\widetilde{S}$, eight fundamentals 
$\hat{Q}$
and bifundamentals $X, \widetilde{X}$
can be described as follows: Two NS5-branes, $(2k'+1)$
NS5'-branes, 
$N_c$- and $N_c'$-D4-branes, and $2N_f$- and 
$2N_f'$-D6-branes, eight D6-branes and $O6^{\pm}$-planes. 
The $N_c$-color D4-branes are suspended between 
the two NS5-branes,
the $N_c'$-color D4-branes are suspended between 
the right NS5-brane and the right NS5'-branes(and their mirrors),
the $N_f$ D6-branes 
are located between the middle NS5'-brane and the right NS5-brane 
and
the $N_f'$ D6-branes 
are located between the right NS5-brane and the right 
NS5'-branes(and their mirrors).

Let us deform this theory
by adding the mass term 
and the quartic term for fundamental quarks.
The former can be achieved by displacing the D6-branes along $\pm v$
direction leading to their coordinates $v = \pm v_{D6}$ \cite{GK98} 
while the latter can be obtained by rotating the D6-branes
\cite{GK0710-1} 
by an angle 
$ \mp \theta$ in $(w,v)$-plane and 
we denote them by $D6_{\mp \theta}$-branes. 
Then, in the electric gauge theory, the general 
deformed superpotential is
given by
\bea
W_{elec} & = & \frac{\alpha}{2} \tr (Q \widetilde{Q})^2 - m \tr Q
\widetilde{Q} + \hat{Q} \widetilde{S} \hat{Q}
 +
 \frac{\alpha'}{2} \tr (Q' \widetilde{Q}')^2 - m' \tr Q'
\widetilde{Q}' \nonu \\
& - & \frac{\beta}{2}   
\tr (A \widetilde{S})^{2}
 +\left[-\frac{\gamma}{2}   \tr (X \widetilde{X})^{k'+1} +
 m_X \tr X
\widetilde{X} \right].
\label{ele}
\eea 
The last three terms are due to the rotation of NS5-branes and NS5'-branes where
$\beta=\tan \omega$ and $\gamma =\tan \omega'$ 
and the relative displacement of D4-branes where the mass $m_X =v_{NS5'}$ 
is the distance in $v$ direction.
The case of $k'=1$ and $\alpha=0$ with $\beta=\gamma=m_X=0$ 
was studied in \cite{Ahn07-4}.
We focus on the case with $k' \geq 2$ and 
$\beta, \gamma,  m_X  \rightarrow 0$.
When we take the Seiberg dual for the gauge group $SU(N_c)$, we put 
$\alpha'=0$ and $m'=0$
and  for 
the Seiberg dual on the gauge group $SU(N_c')$ we take 
$\alpha=0$ and $m=0$.

Let us summarize the ${\cal N}=1$ supersymmetric electric brane
configuration with superpotential 
(\ref{ele}) 
in type IIA string theory as follows and draw this in
Figure 20:

$\bullet$
Two NS5-branes in $(012345)$ directions  with $w=0$

$\bullet$ 
$(2k'+1)$ NS5'-branes in  $(012389)$ directions $v=0$

$\bullet$
$N_f$ $D6_{\pm \theta}$-branes in (01237)
directions and
two other directions in $(v,w)$-plane

$\bullet$
$N_f'$ $D6_{\pm \theta'}$-branes in (01237) directions
  and
two other directions in $(v,w)$-plane

$\bullet$
Eight D6-branes in (0123789) directions 

$\bullet$
$N_c$- and $N_c'$-color D4-branes in $(01236)$ directions  with $v=0=w$ 

$\bullet$ $O6^{\pm}$-planes in (0123789) directions with $x^6=0=v$ 

\begin{figure}[ht]
   \epsfxsize=4.5in 
\centerline{\epsffile{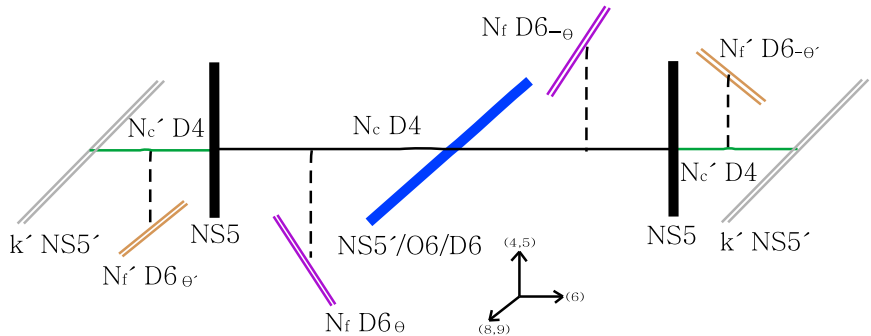}}
   \caption[FIG. \arabic{figure}.]{ 
The  ${\cal N}=1$ supersymmetric 
electric brane configuration for the gauge group $SU(N_c) \times SU(N_c')$ 
with an antisymmetric tensor $A, \widetilde{S}$, 
bifundamentals $X, \widetilde{X}$, fundamentals $Q, \widetilde{Q},
Q', \widetilde{Q}'$ and fundamentals $\hat{Q}$. 
Note that there are multiple $2k'$ $NS5'$-branes.
A 
rotation of $N_f$ D6-branes in $(w,v)$-plane
corresponds to 
a quartic term for the fundamentals $Q, \widetilde{Q}$ while 
a displacement of $N_f$ D6-branes in $ \pm v$ direction corresponds to a
mass term for the fundamentals $Q, \widetilde{Q}$.
}
\end{figure}

\subsection{Magnetic theory for $SU(N_c)$}

After the left NS5-branes and $D6_{\theta}$-branes and 
the right NS5-branes and $D6_{-\theta}$-branes 
are exchanged each other, we arrive at the Figure 21A.
The number of colors of dual magnetic theory  is given by \cite{Ahn07-4} 
$
\widetilde{N}_c = 2(N_f+N_c')-N_c+4$.

\begin{figure}[ht]
   \epsfxsize=4.5in 
\centerline{\epsffile{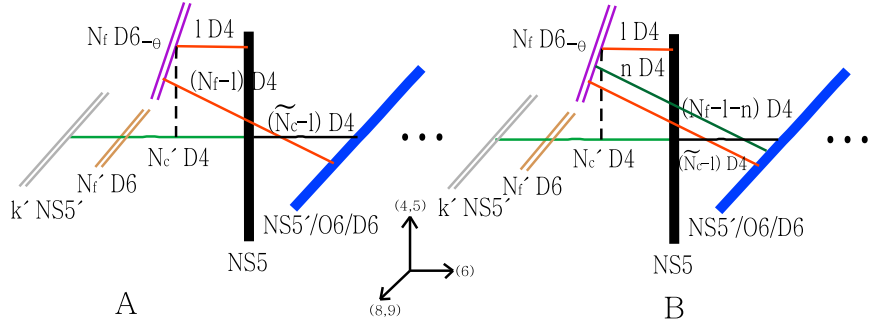}}
   \caption[FIG. \arabic{figure}.]{ 
The  ${\cal N}=1$ supersymmetric
magnetic brane configuration corresponding to Figure 20 with 
a misalignment between D4-branes when the gravitational potential of
the NS5-brane is ignored(21A) and nonsupersymmetric brane configuration
when  the gravitational potential of
the NS5-brane is considered(21B). 
The $N_f$ flavor D4-branes connecting between
$D6_{-\theta}$-branes and NS5'-brane are splitting into $(N_f-l)$- and
$l$- D4-branes(21A). Further $n$- D4-branes among 
$(N_f-l)$- D4-branes are moved to the NS5-brane(22B). 
}
\end{figure}

The low energy theory on the color D4-branes 
has $SU(\widetilde{N}_c) \times SU(N_c')$ gauge group and  
$N_f$-fundamental dual quarks $q, \widetilde{q}$
coming from 4-4 strings connecting between the color $\widetilde{N}_c$ D4-branes and
$N_f$ flavor D4-branes as well as $Q', \widetilde{Q}', a,
\widetilde{s}, \hat{q}, Y, \widetilde{Y}$ and gauge singlets.
Moreover, a magnetic meson field $M \equiv Q \widetilde{Q}$
is $N_f \times N_f$ matrix and comes from 
4-4 strings of $N_f$ flavor D4-branes.
Then the magnetic superpotential with the limit 
$\beta, \gamma, m_X
\rightarrow 0$ 
is given by  
\bea
W_{dual} = \left[ \frac{1}{\Lambda} 
M q \widetilde{s} a \widetilde{q} + \hat{q} \widetilde{s} \hat{q} 
+ Y \widetilde{F}' \widetilde{q} + 
\widetilde{Y} q F' + \Phi' Y \widetilde{Y} + \widetilde{M} \hat{q}
\widetilde{q} 
\right] 
+\frac{\alpha}{2} \tr M^2 - m M.
\label{dualWone}
\eea
The case with $k'=1$ and $\alpha=0$ was studied in \cite{Ahn07-4} as
mentioned before.
Although the superpotential (\ref{dualWone}) 
does not depend on the multiplicity  $k'$ of
NS5'-branes in our particular limit, 
the difference from the previous result  appears as
1) nonzero $\alpha$ in (\ref{dualWone}) and 2) multiple NS5'-branes
in Figure 21A.
Here other meson fields are given by 
$\Phi' \equiv X \widetilde{X},  
F' \equiv \widetilde{X} Q, \widetilde{M} \equiv \hat{Q} \widetilde{Q}$ and 
$\widetilde{F}' \equiv X \widetilde{Q}$ \cite{Ahn07-4}.

For the supersymmetric vacua, one can compute the F-term equations for
this superpotential (\ref{dualWone}) 
and the expectation values for $M$ and $q \widetilde{s} a \widetilde{q}$ 
are obtained.
The F-term equations are almost the same as the one in \cite{Ahn07-4}
and the derivative of (\ref{dualWone}) with respect 
to the meson field $M$ has $\alpha$ dependent term.
The vacuum expectation values for $Y, \widetilde{Y}, \hat{q},
\widetilde{M},
F'$ and
$\widetilde{F}'$
vanish as in \cite{Ahn07-4}.  

$\bullet$ Coincident $N_f$ $D6_{-\theta}$-branes 

The theory has many nonsupersymmetric meta-stable ground states and 
when we rescale the meson field as
$
M = h \Lambda \Phi$ as before,
then the Kahler potential for $\Phi$ is canonical and the magnetic
quarks are canonical near the origin of field space \cite{ISS}.
Then the magnetic superpotential can be written as
\bea
W_{mag} = h \Phi  q  \widetilde{s} a \widetilde{q} 
 +  
\frac{h^2 \mu_{\phi}}{2} \tr \Phi^2- h \mu^2 \tr \Phi +
\hat{q} \widetilde{s} \hat{q}+ Y F q' + 
\widetilde{Y} \widetilde{q}' \widetilde{F}' 
+ \Phi' Y \widetilde{Y} + \widetilde{M} \hat{q}
\widetilde{q} 
\nonu
\eea
where
$
\mu^2 = m \Lambda$ and  
$\mu_{\phi} = \alpha \Lambda^2$.
Now one splits 
the $(N_f-l) \times (N_f-l)$
block  at the lower right corner of $h\Phi$ and $q 
\widetilde{s} a \widetilde{q}$ 
into blocks of 
size $n$ and $(N_f-l-n)$ as follows \cite{GK0710}:
\bea
h\Phi = \left(
\begin{array}{ccc}
0_l & 0 & 0  \\
0 & h \Phi_n & 0 \\
0 & 0 & \frac{\mu^2}{\mu_{\phi}} {\bf 1}_{N_f-l-n}
\end{array}
\right), \qquad
q \widetilde{s} a  \widetilde{q} = \left(
\begin{array}{ccc}
\mu^2 {\bf 1}_l & 0 & 0  \\
0 & { \varphi} \widetilde{\beta} \gamma \widetilde{\varphi}  &  0 \\
0 & 0 & 0_{N_f-l-n}
\end{array}
\right).
\nonu
\eea
Here $\varphi$ and $\widetilde{\varphi}$ are $n \times (\widetilde{N}_c-l)$
dimensional matrices and correspond to $n$ flavors of fundamentals of
the gauge group $SU(\widetilde{N}_c-l)$ which is unbroken.
The $\Phi_n$ and ${ \varphi} \widetilde{\beta}
\gamma \widetilde{\varphi}$
are $n \times n$ matrices.
The supersymmetric ground state corresponds to
$h\Phi_n= \frac{\mu^2}{\mu_{\phi}} {\bf 1}_{n}$ and $ 
\varphi \widetilde{\beta} =0= \gamma \widetilde{\varphi}$. 

Now the full one loop potential 
takes the form
\bea
\frac{V}{|h|^2}  =  
|\Phi_n  \varphi \widetilde{\beta} +  \widetilde{Y} \widetilde{F}|^2   
+  |\Phi_n \gamma \widetilde{\varphi} + F Y|^2
  +  
| \varphi  \widetilde{\beta} \gamma \widetilde{\varphi}-\mu^2 {\bf 1}_{n} + 
h \mu_{\phi} \Phi_n|^2 + b |h \mu|^2 \tr \Phi_n^{\dagger} \Phi_n, 
\nonu
\eea
where $b = \frac{(\ln 4-1)}{8\pi^2} \widetilde{N}_c$ and we do not
write down $\Phi_n$ or $\Phi_n^{\dagger}$-independent terms.
Differentiating this potential with respect to 
$\Phi_n^{\dagger}$ and putting $\varphi \widetilde{\beta}=0=
\gamma \widetilde{\varphi}$, one obtains
\bea
h \Phi_n 
\simeq \frac{ \mu_{\phi}}{b }
{\bf 1}_n \qquad \mbox{or} \qquad
M_n \simeq \frac{\alpha \Lambda^3}{\widetilde{N}_c} {\bf 1}_{n}
\nonu
\eea
corresponding to the $w$ coordinates of $n$ curved flavor D4-branes between 
the $D6_{-\theta}$-branes and the NS5'-branes.

$\bullet$ Non-coincident $N_f$ $D6_{-\theta}$-branes 

These  non-coincident $D6_{-\theta}$-branes can be obtained by taking those
quark masses being unequal.

\subsection{Magnetic theory for $SU(N_c')$}

After we move the right NS5-brane to the right all the way 
past the right 
NS5'-branes(and its mirrors to the left), we arrive at the Figure 22A.
The linking number of NS5-brane from Figure 22A
is 
$
l_m = \frac{N_f'}{2} -\widetilde{N}_c$.
On the other hand, the linking number of NS5-brane from Figure 21
is
$
l_e = -\frac{N_f'}{2} + N_c -N_c'$. 
From these two relations, one obtains
the number of colors of dual magnetic theory
$
\widetilde{N}_c' = N_f' + N_c - N_c'$.

\begin{figure}[ht]
   \epsfxsize=5.0in 
\centerline{\epsffile{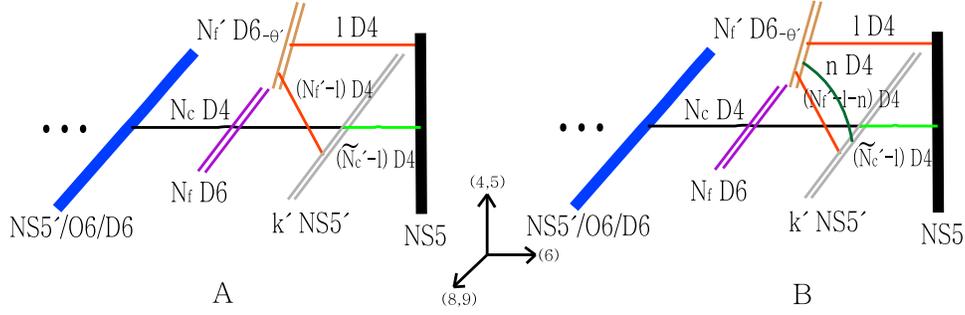}}
   \caption[FIG. \arabic{figure}.]{ 
The  ${\cal N}=1$ supersymmetric
magnetic brane configuration corresponding to Figure 20 with 
a misalignment between D4-branes when the gravitational potential of
the NS5-brane is ignored(22A) and nonsupersymmetric brane configuration
when  the gravitational potential of
the NS5-brane is considered(22B).
Note that there are multiple $2k'$ $NS5'$-branes. 
The $N_f'$ flavor D4-branes connecting between
$D6_{-\theta'}$-branes and NS5'-brane are splitting into $(N_f'-l)$- and
$l$- D4-branes(22A). Further $n$- D4-branes among 
$(N_f'-l)$- D4-branes are moved to the NS5-brane(22B). 
}
\end{figure}

The low energy theory on the color D4-branes 
has $SU(N_c) \times SU(\widetilde{N}_c')$ gauge group and  
$N_f'$-fundamental dual quarks $q', \widetilde{q}'$
coming from 4-4 strings connecting between the color $\widetilde{N}_c'$ D4-branes and
$N_f'$ flavor D4-branes as well as $Q, \widetilde{Q}, A,
\widetilde{S}, \hat{Q}, Y, \widetilde{Y}$ and gauge singlets.
Moreover, a single magnetic meson field $M \equiv Q' \widetilde{Q}'$
is $N_f' \times N_f'$ matrix and comes from 
4-4 strings of $N_f'$ flavor D4-branes.
Then the magnetic superpotential with the limit $\beta, \gamma, m_X
\rightarrow 0$ 
is given by  
\bea
W_{dual} = \left[ \frac{1}{\Lambda} M q' \widetilde{q}'    + 
Y F q' +
\widetilde{Y} \widetilde{q}' \widetilde{F} + \Phi' Y \widetilde{Y} \right] 
+\frac{\alpha}{2} \tr M^2 - m' M.
\label{dual1fin}
\eea
The case with $k'=1$ and $\alpha=0$ was studied in \cite{Ahn07-4}.
Although the superpotential (\ref{dual1fin}) 
does not depend on the multiplicity  $k'$ of
NS5'-branes in our particular limit, 
the difference from the previous result of \cite{Ahn07-4} appears as
two things: nonzero $\alpha$ in (\ref{dual1fin}) and multiple NS5'-branes
in Figure 22A.
Here other meson fields are given by 
$\Phi' \equiv X \widetilde{X},  
F \equiv X Q'$ and 
$\widetilde{F} \equiv \widetilde{X} \widetilde{Q}'$ \cite{Ahn07-4}.

For the supersymmetric vacua, one can compute the F-term equations for
this superpotential (\ref{dual1fin}) 
and the expectation values for $M$ and $q' \widetilde{q}'$ 
are obtained.
The F-term equations are almost the same as the one in \cite{Ahn07-4}
and the derivative of (\ref{dual1fin}) with respect 
to the meson field $M$ has $\alpha$ dependent term.
The vacuum expectation values for $Y, \widetilde{Y}, F$ and
$\widetilde{F}$
vanish as in \cite{Ahn07-4}.  

$\bullet$ Coincident $N_f'$ $D6_{-\theta'}$-branes  

One obtains the stable point (\ref{sec5vac}) with appropriate
replacement of $N_f$ $D6_{-\theta}$-branes by the corresponding $N_f'$
$D6_{-\theta'}$-branes.
This gives the $w$ coordinates of $n$ curved flavor D4-branes between 
the $D6_{-\theta'}$-branes and the NS5'-branes in Figure 22B.

$\bullet$ Non-coincident $N_f'$ $D6_{-\theta'}$-branes and $k'$ NS5'-branes 

The local nonzero stable point arises as (\ref{sec5vac1}).

\section{
Conclusions and outlook }

As mentioned in the abstract, let us summarize 
the new features we have obtained from the 
meta-stable brane configurations.

Compared with the ones of \cite{GK0710,GK0710-1}, 
the Figure 2 contains the multiple $k'$ NS5'-branes and the presence
of
$\widetilde{N}_c$ flavor D4-branes connecting $D6_{-\theta}$-branes
and NS5'-branes. 
Similarly, its symplectic version, characterized by the Figure 4, contains
the multiple $(2k'+1)$ NS5'-branes and the presence
of
$\widetilde{N}_c$ flavor D4-branes connecting $D6_{-\theta}$-branes
and NS5'-branes(and its mirrors).  
Compared with the one of \cite{Ahn07-11}, 
the Figure 6 has the multiple $k'$ NS5'-branes and the different 
value of $\widetilde{N}_c$ which has $k'$-dependence. 
The Figures 8, 10, 12 have multiple NS5'-branes and rotated D6-branes,
compared with \cite{Ahn07-3}.

Compared with \cite{Ahn07-5}, the Figure 14 has multiple rotated NS5-branes
and multiple NS5'-branes while the Figure 16 has multiple rotated NS5'-branes.
The Figure 18 contains the multiple $k'$ NS5'-branes and the different 
value of $\widetilde{N}_c$ which has $k'$-dependence while the Figure
19 has only rotated D6-branes, compared with the previous result 
in \cite{Ahn07-4}. 
Finally, the Figure 21 has only rotated D6-branes and the Figure 22 
multiple NS5'-branes and rotated D6-branes,
compared with \cite{Ahn07-4}.

We make some comments for the future directions along the line of
meta-stable brane construction.

$\bullet$ As we mentioned,  the construction for 
possible multiple outer NS5-branes on
the gauge theory \cite{ILS} with the antisymm. and
conj. symm. as well as fundamentals is an open problem in the context
of the analysis of supersymmetric ground states and nonsupersymmetric
ground states also. 

$\bullet$ 
Are there any variants of sections 2 and 3 by 
$k$ NS5-branes and a single NS5'-brane?
When $k=2$, 
the gauge theory analysis for unitary group was given in \cite{AGM}.
It would be interesting to study whether there exist nonsupersymmetric 
meta-stable vacua 
and if not, how one can think of the possible 
deformations in the superpotential to make them stabilize 
at one-loop?

$\bullet$ When there are multiple $k'$ middle NS5-branes, which is the
same number as the one of NS5'-branes, in section 4,
how one can analyze the meta-stable vacuum? For the supersymmetric
brane configuration corresponding to the gauge theory 
\cite{Brodie,BS,Mazz}, some
of the analysis was done in \cite{BHKL} where there are the higher order term
for an adjoint field in the superpotential 
and the interaction terms between an adjoint and
quarks and symmetric tensor.   

$\bullet$ Similarly, when there are multiple $k'$ middle NS5-branes in
section 5(there are $(2k'+1)$ middle NS5-branes in section 6), 
how one can analyze the meta-stable vacuum? 
For the supersymmetric
brane configuration \cite{BH} 
corresponding to this gauge theory \cite{BS} which
has extra two adjoint fields, the superpotential has more general
form.
 
$\bullet$ It is possible to  consider the case where 
there are multiple $k$ NS5-branes and a single NS5'-brane(and their
mirrors) in section 7.
It would be interesting to study whether there exist nonsupersymmetric 
meta-stable vacua. 

$\bullet$ As mentioned before, 
when there are multiple $k'$ middle NS5-branes in section 8, 
the analysis for the supersymmetric
brane configuration \cite{BHKL} explains the gauge theory 
result \cite{BS}. It is an open problem to find whether 
there exist nonsupersymmetric 
meta-stable vacua. 

$\bullet$ Also one can think of  the case when 
there are multiple $k$ NS5-branes in section 9.  
It would be interesting to find out 
the supersymmetric ground states and nonsupersymmetric
ground states.

$\bullet$ When there are multiple $k'$ middle NS5-branes in section
10, how one can analyze the meta-stable vacuum? 
It is known that 
the supersymmetric
brane configuration \cite{BHKL} describes the gauge theory \cite{BS}.
On the other hand, one can consider the case where 
there exist multiple $k$ NS5-branes with a single NS5'-brane  or
combination of both multiple $k'$ middle NS5-branes and 
multiple $k$ NS5-branes in section 10.
For the dual of second gauge group $SU(N_c')$, we treated only 
$k'=1$ NS5'-brane but it is also possible to consider the general 
$k'$ case where there are many meson fields.

$\bullet$ Also when there are multiple $k$  NS5-branes with 
a single NS5'-brane, 
or when there are multiple $k$ NS5-branes and  multiple $k'$
NS5'-branes in section 11, how the meta-stable vacua as well as
supersymmetric ones appear?  

\vspace{.7cm}

\centerline{\bf Acknowledgments}

This work was supported by grant No.
R01-2006-000-10965-0 from the Basic Research Program of the Korea
Science \& Engineering Foundation.  
I would like to thank KIAS(Korea Institute for 
Advanced Study) for hospitality  where
this work was undertaken.

\end{document}